\documentclass[usenatbib]{mnras}

\usepackage[T1]{fontenc}

\DeclareRobustCommand{\VAN}[3]{#2}
\let\VANthebibliography\thebibliography
\def\thebibliography{\DeclareRobustCommand{\VAN}[3]{##3}\VANthebibliography}

\usepackage{graphicx}	
\usepackage[flushleft]{threeparttable}
\usepackage{booktabs}
\usepackage{amsmath}
\usepackage{subfigure}
\usepackage{pdflscape}
\usepackage{array}
\usepackage{tabularx}
\usepackage{verbatim}
\usepackage{colortbl}
\usepackage{hyperref}
\usepackage{float}
\usepackage{adjustbox}
\usepackage{multirow}

\def \teff {$T_{\mathrm{eff}}$}

\newcommand{\mytilde}{{\raise.17ex\hbox{$\scriptstyle\sim$}}}

\providecommand{\teff}{\ensuremath{T_{\rm eff}}}

\usepackage{txfonts}

\title[The TOI-1052 planetary system]{Discovery and characterisation of two Neptune-mass planets orbiting HD 212729 with TESS}
   
\author[D. J. Armstrong et. al.]{
David J. Armstrong$^{1,2}$,\thanks{E-mail: d.j.armstrong@warwick.ac.uk},
Ares Osborn$^{1,2}$,
Vardan Adibekyan$^{3}$,
Elisa~Delgado-Mena$^{3}$,
Saeed~Hojjatpanah$^{4}$,
\newauthor
Steve~B.~Howell$^{5}$,
Sergio~Hoyer$^{4}$,
Henrik Knierim$^{6}$,
S\'ergio G. Sousa$^{3}$,
Keivan G. Stassun$^{7}$, 
\newauthor
Dimitri Veras$^{1,2,8}$,
David~R.~Anderson$^{1,2}$,
Daniel Bayliss$^{1,2}$,
Fran\c{c}ois Bouchy$^{9}$,
Christopher~J.~Burke$^{10}$,
\newauthor
Jessie~L.~Christiansen$^{11}$,
Xavier Dumusque$^{9}$,
Marcelo Aron Fetzner Keniger$^{1,2}$,
Andreas Hadjigeorghiou$^{1,2}$,
\newauthor
Faith Hawthorn$^{1,2}$,
Ravit Helled$^{6}$,
Jon~M.~Jenkins$^{5}$,
David W.\ Latham$^{12}$,
\newauthor
Jorge Lillo-Box$^{13}$,
Louise D.\ Nielsen$^{14}$,
Hugh P.\ Osborn$^{15}$,
Jos\'e Rodrigues$^{3}$,
David Rodriguez$^{16}$,
\newauthor
Nuno C. Santos$^{3,17}$,
Sara Seager$^{10,18,19}$,
Paul A. Str{\o}m$^{1,2}$,
Guillermo Torres$^{12}$,
Joseph D. Twicken$^{20,5}$,
\newauthor
Stephane Udry$^{9}$,
Peter J. Wheatley$^{1,2}$,
Joshua N.\ Winn$^{21}$
\\
$^{1}$Department of Physics, University of Warwick, Coventry CV4 7AL, UK\\
$^{2}$Centre for Exoplanets and Habitability, University of Warwick, Gibbet Hill Road, Coventry CV4 7AL, UK\\
$^{3}$Instituto de Astrof\'isica e Ci\^encias do Espa\c{c}o, Universidade do Porto, CAUP, Rua das Estrelas, 4150-762 Porto, Portugal\\
$^{4}$Laboratoire d'Astrophysique de Marseille, Pole de l'Etoile Site de Chateau-Gombert, 38 Rue Frederic Joliot-Curie, Marseille, FR 13338\\
$^{5}$NASA Ames Research Center, Moffett Field, CA 94035, USA\\
$^{6}$Institute for Computational Science, Center for Theoretical Astrophysics \& Cosmology, University of Zurich, Winterthurerstr. 90, 8057 Zurich, Switzerland\\
$^{7}$Department of Physics and Astronomy, Vanderbilt University, Nashville, TN 37235, USA\\
$^{8}$Centre for Space Domain Awareness, University of Warwick, Gibbet Hill Road, Coventry CV4 7AL, UK \\
$^{9}$Department of Astronomy of the University of Geneva, Geneva Observatory, Chemin Pegasi 51, 1290 Versoix, Switzerland\\
$^{10}$Department of Physics and Kavli Institute for Astrophysics and Space Research, Massachusetts Institute of Technology, Cambridge, MA 02139, USA\\
$^{11}$Caltech/IPAC, Mail Code 100-22, 1200 E. California Blvd. Pasadena, CA 91125\\
$^{12}$Center for Astrophysics \textbar \ Harvard \& Smithsonian, 60 Garden Street, Cambridge, MA 02138, USA\\
$^{13}$Centro de Astrobiolog\'ia (CAB, CSIC-INTA), Depto. de Astrof\'isica, ESAC campus, 28692, Villanueva de la Ca\~nada (Madrid), Spain\\
$^{14}$European Southern Observatory, Karl-Schwarzschild-Stra{\ss}e 2, 85748 Garching bei M{\"u}nchen, Germany\\
$^{15}$Physikalisches Institut, University of Bern, Gesellsschaftstrasse 6, 3012 Bern, Switzerland\\
$^{16}$Space Telescope Science Institute, 3700 San Martin Drive, Baltimore, MD, 21218, USA\\
$^{17}$Departamento de F\'isica e Astronomia, Faculdade de Ci\^encias, Universidade do Porto, Rua do Campo Alegre, Porto, Portugal\\
$^{18}$Department of Earth, Atmospheric and Planetary Sciences, Massachusetts Institute of Technology, Cambridge, MA 02139, USA\\
$^{19}$Department of Aeronautics and Astronautics, MIT, 77 Massachusetts Avenue, Cambridge, MA 02139, USA\\
$^{20}$SETI Institute, Mountain View, CA 94043 USA\\
$^{21}$Department of Astrophysical Sciences, Princeton University, Princeton, NJ 08544, USA
    }

\date{Accepted XXX. Received YYY; in original form ZZZ}

\pubyear{2023}

\begin{document} 
\label{firstpage}
\pagerange{\pageref{firstpage}--\pageref{lastpage}}
\maketitle

\begin{abstract}
  We report the discovery of two exoplanets orbiting around HD 212729 (TOI\,1052, TIC 317060587), a $T_{\rm eff}=6146$K star with V=9.51 observed by \textit{TESS} in Sectors 1 and 13. One exoplanet, TOI-1052b, is Neptune-mass and transits the star, and an additional planet TOI-1052c is observed in radial velocities but not seen to transit. We confirm the planetary nature of TOI-1052b using precise radial velocity observations from HARPS and determined its parameters in a joint RV and photometry analysis. TOI-1052b has a radius of $2.87^{+0.29}_{-0.24}$ R$_{\oplus}$, a mass of $16.9\pm 1.7$ M$_{\oplus}$, and an orbital period of 9.14 days. TOI-1052c does not show any transits in the \textit{TESS} data, and has a minimum mass of $34.3^{+4.1}_{-3.7}$ M$_{\oplus}$ and an orbital period of 35.8 days, placing it just interior to the 4:1 mean motion resonance. Both planets are best fit by relatively high but only marginally significant eccentricities of $0.18^{+0.09}_{-0.07}$ for planet b and $0.24^{+0.09}_{-0.08}$ for planet c. We perform a dynamical analysis and internal structure model of the planets as well as deriving stellar parameters and chemical abundances. The mean density of TOI-1052b is $3.9^{+1.7}_{-1.3}$ g cm$^{-3}$ consistent with an internal structure similar to Neptune. A nearby star is observed in Gaia DR3 with the same distance and proper motion as TOI-1052, at a sky projected separation of \mytilde1500AU, making this a potential wide binary star system.
\end{abstract}

\begin{keywords}
exoplanets -- planets and satellites: detection -- planets and satellites: individual: (TOI-1052, TIC 317060587)
\end{keywords}

\section{Introduction}

The Transiting Exoplanet Survey Satellite (\textit{TESS}), launched in April 2018, has discovered more than 6000 exoplanet candidates and more than 300 confirmed exoplanets to date \cite[][]{Ricker-15}, \cite[see NASA Exoplanet Archive,][]{Akeson-13}. \textit{TESS} has observed nearly the entire sky, in a series of 27-day sectors, with many stars observed for two or more sectors.

This large number of candidates is made up of both real exoplanets and false positives. High resolution and stable spectroscopy for radial velocity measurements is essential not only to confirm the candidate exoplanets but also to measure the mass of the planet precisely. Knowing both mass and radius of a planet provides us an estimation of the bulk density, internal structure and planetary composition. High resolution spectroscopy allows us to fully derive stellar parameters and chemical abundances allowing studies of planetary origin, formation and evolution \citep[][]{Alibert-10,Alibert-15}. 

The discovery of thousands of exoplanets over recent years has revealed a diverse population of exoplanets ranging from Earth-mass to Jupiter-mass, which also can be categorised in various sub-types including hot-Jupiters and super-Earths. A less well known group is the Neptunian exoplanets, raising questions about their overall occurrence, and an apparent lack of short period Neptune-mass exoplanets - the so-called Neptunian desert \citep[][]{Szabo-11,Mazeh2016} which could be due to tidal disruption and photoevaporation \citep[][]{Beauge-13,Mazeh2016}, combined with potential migration of evaporating planets \citep{Boue2012}. Different programs including our NCORES and NOMADS HARPS large programs aim at a more detailed investigation of Neptunian planets, and have successfully discovered several such exoplanets \citep[e.g. ][]{Nielsen2020,Otegi2021,SHoyer2021}. The number of discoveries in this parameter space remains low and more exoplanets are needed to better understand the population.

\begin{table}
\caption{\label{tab:stellar} Stellar parameters.}
\resizebox{\columnwidth}{!}{%
	\begin{tabular}{lcc}
	\hline\hline
	\noalign{\smallskip}
	Parameter	&	Value	&	Source\\

	\hline
    \noalign{\smallskip}
    \noalign{\smallskip}
    \multicolumn{3}{l}{\underline{Identifying Information}}\\
    \noalign{\smallskip}
    \noalign{\smallskip}
    Identifier & HD 212729 & \\
    TOI & TOI-1052 &  \textit{TESS} \\
    TIC  ID & 317060587 &  \textit{TESS} \\
    2MASS ID	&  22300272-7538476	& 2MASS \\
    \multicolumn{2}{l}{Gaia ID ~~~~~~~~~~~~~~~~~~~~~~~~~~~6357524189130820992}	& Gaia DR3 \\

    \\
    \multicolumn{3}{l}{\underline{Astrometric Parameters}}\\
    \noalign{\smallskip}
    \noalign{\smallskip}
    R.A. (J2000, deg)		&	337.51023851341 	&  Gaia DR3	\\
	Dec	 (J2000, deg)		&	 -75.64656089247	&  Gaia DR3	\\
    Parallax  (mas) & 7.74 $\pm$0.01 & Gaia DR3\\
    Distance  (pc) & 128.7$\pm$0.2 & Gaia DR3\\
    \\
    \multicolumn{3}{l}{\underline{Photometric Parameters}}\\
    \noalign{\smallskip}
    \noalign{\smallskip}
	B 		&10.09 $\pm$ 0.04  	& Tycho \\
	V 		&9.51 $\pm$ 0.02	&Tycho\\
	T 	    & 9.02 $\pm$ 0.01	& \textit{TESS}\\
    G 		& 9.40 $\pm$ 0.01	&Gaia DR3\\
    J 		& 8.56  $\pm$ 0.02	&2MASS\\
   	H 		&  8.25 $\pm$ 0.04	&2MASS\\
	K      & 8.25$\pm$ 0.03	&2MASS\\
    W1  & 8.160 $\pm$ 0.023	&WISE\\
    W2  & 8.255 $\pm$ 0.021	&WISE\\
    W3  & 8.190  $\pm$ 0.020 & WISE\\
    W4  & 8.119 $\pm$ 0.174 & WISE \\
    
    \\
    
    \multicolumn{3}{l}{\underline{Abundances}}\\
    \noalign{\smallskip}
    \noalign{\smallskip}
	$\mathrm{\left[Fe/H\right](dex)}$  & 0.140 $\pm$ 0.013 &Sec. \ref{sec:star} \\
	$\mathrm{\left[MgI/H\right](dex)}$  & 0.08 $\pm$ 0.04 &Sec. \ref{sec:star} \\
	$\mathrm{\left[AlI/H\right](dex)}$  & 0.10 $\pm$ 0.02 &Sec. \ref{sec:star} \\
	$\mathrm{\left[SiI/H\right](dex)}$  & 0.11 $\pm$ 0.04 &Sec. \ref{sec:star} \\
$\mathrm{\left[TiI/H\right](dex)}$  & 0.12 $\pm$ 0.03 &Sec. \ref{sec:star} \\
	$\mathrm{\left[NiI/H\right](dex)}$  & 0.13 $\pm$ 0.03 &Sec. \ref{sec:star} \\

\\

    \\

  \multicolumn{3}{l}{\underline{Bulk Parameters}}\\
    \noalign{\smallskip}
    \noalign{\smallskip}
    Mass ($M_{\odot}$)                   & 1.204 $\pm$ 0.025  & Sec. \ref{sec:star} (\texttt{PARAM})\\
    Radius ($R_{\odot}$)                 & 1.264 $\pm$ 0.033  & Sec. \ref{sec:star} (\texttt{PARAM})\\
    \teff\,(K)                           & 6146 $\pm$ 62    & Sec. \ref{sec:star} \\
    log g (cm\,s$^{-2}$)                 & 4.30 $\pm$ 0.02  & Sec. \ref{sec:star} (Gaia)\\
    $\log g$ (cm\,s$^{-2}$)              & $4.39  \pm 0.11$ & Sec. \ref{sec:star} (spec)\\
    $v_\mathrm{mic}$ (km~s$^{-1}$)        & $1.28  \pm 0.02$              & Sec. \ref{sec:star} \\ 
 	{\it v}\,sin\,{\it i} (km\,s$^{-1}$) & $5.0 \pm 0.9$	& Sec. \ref{sec:star} \\
 	P$_{rot}$/\,sin\,{\it i} (d)                       & $12.8 \pm 2.3$	&Sec. \ref{sec:star} \\
	Age	(Gyrs)                           & 2.3 $\pm$ 1.0	&Sec. \ref{sec:star} (\texttt{PARAM}) \\
	 mean $log(R' _{HK})$   & -5.32 $\pm$ 0.02 & HARPS \\ 

 	\noalign{\smallskip}
	\hline
 	\noalign{\smallskip}
    \end{tabular}}
    
 Sources: \textit{TESS} \citep{Stassun2019TIC};2MASS \citep{2MASS};Tycho \citep{Tycho}; WISE \citep{WISE}; and Gaia \citep{GaiaMission2016,GaiaDR32022} 
\end{table}       

In this paper we report the discovery and confirmation of TOI-1052 b, a Neptune-like planet orbiting HD 212729, a G0 high proper motion southern star with a visual magnitude of 9.51 (\textit{TESS} mag 
 of 9.02) and a non-transiting additional planet. We use high precision radial
velocity (RV) measurements from the High Accuracy Radial
velocity Planet Searcher spectrograph \cite[HARPS, ][]{Pepe-02}
mounted at the ESO La Silla 3.6 m telescope, in the framework of
the NCORES program \cite[e.g.][]{Nielsen2020,Armstrong2020}. Simultaneous analysis of the HARPS RV measurements using high resolution spectroscopy and \textit{TESS} photometry enable us to confirm the nature of the planets as well as determine the stellar parameters of the host star.

The paper is organised as follows: the observations and data of the system are described in Sect. \ref{sect:obs}. The stellar parameters and signal analysis are described in Sect. \ref{sec:star}. In Sect. \ref{sect:model} we describe the joint model and the discussion is presented in Sect. \ref{sect:discuss}.

\vspace{5mm}

\section{Observations}
\label{sect:obs}
\subsection{\textit{TESS} photometry}


\textit{TESS} observed TOI\,1052 (TIC 317060587) in Sectors 1 and 13, obtaining data from 25 July to 22 August 2018 and from 19 June to 17 July 2019. The data were reduced in the TESS Science Processing Operations Center \citep[SPOC,][]{Jenkins-16} pipeline. This pipeline is adapted from Kepler mission pipeline at NASA Ames Research Center \citep{Jenkins2010a}. Transit events with 9.1 d orbital period were detected in the SPOC search of the two-minute cadence light curve for Sector 1 on 28 Aug 2018 and for Sector 13 on 27 July 2019 with an adaptive, noise-compensating matched filter \citep{Jenkins2002,Jenkins2010b,Jenkins2020}. A limb-darkened transit model was fitted \citep{Li-19} and a suite of diagnostic tests were conducted to assess the planetary nature of the signal \citep{Twicken-18}. The TESS Science Office (TSO) reviewed the SPOC data validation reports and issued an alert for TOI 1052.01 following the Sector 13 transit search on 16 August 2019 \citep{Guerrero2021}. In a search of the combined data from both sectors, the SPOC reported a signal-to-noise ratio (SNR) of 10.62 and transit depth value of 358.4 +- 33.7 parts-per-million (ppm) for the 9.14 d period transit event. The transit signal passed all diagnostic tests, and the source was localized within 2.12 +/- 4.55’’ of the target star.

\begin{figure*}
  \begin{center}
    \includegraphics[scale=0.5]{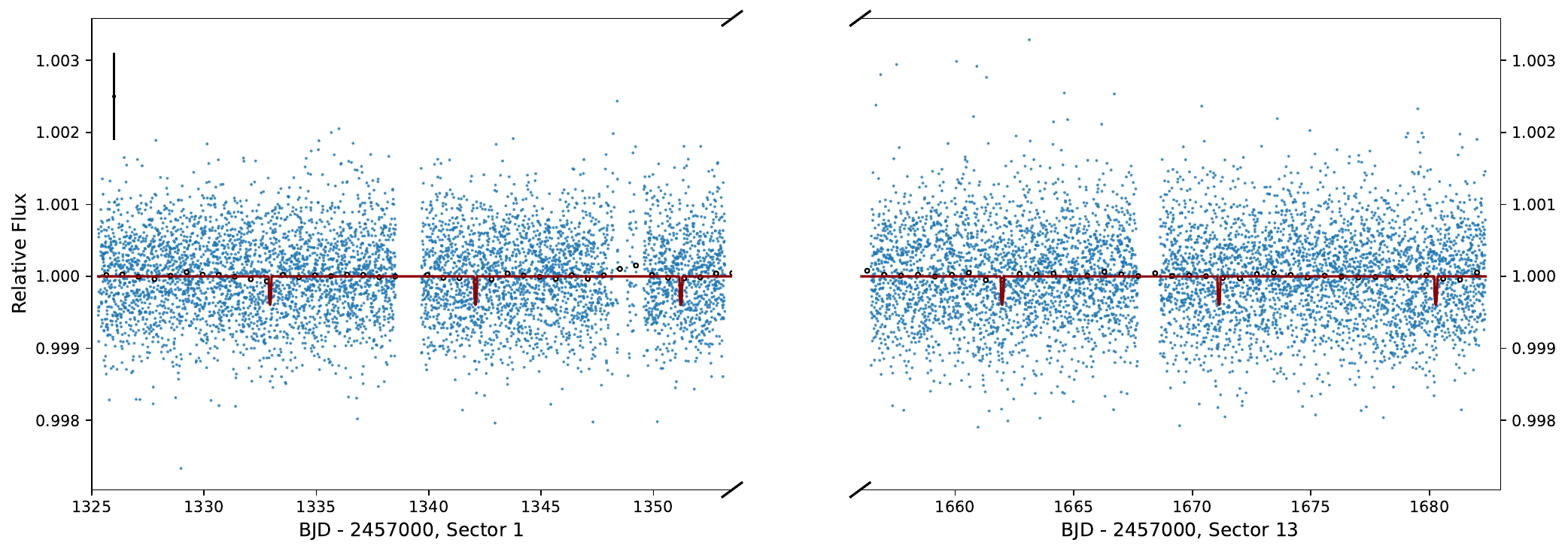}
    \caption{Full \textit{TESS} PDCSAP lightcurve of TOI-1052 at 2-minute cadence with the best fit model overplotted in red. A typical error bar is shown in the top left. Binned datapoints of width 0.7d are shown.\label{fig:lcall}}
  \end{center}
\end{figure*}

\begin{figure}
  \begin{center}
    \includegraphics[scale=0.5]{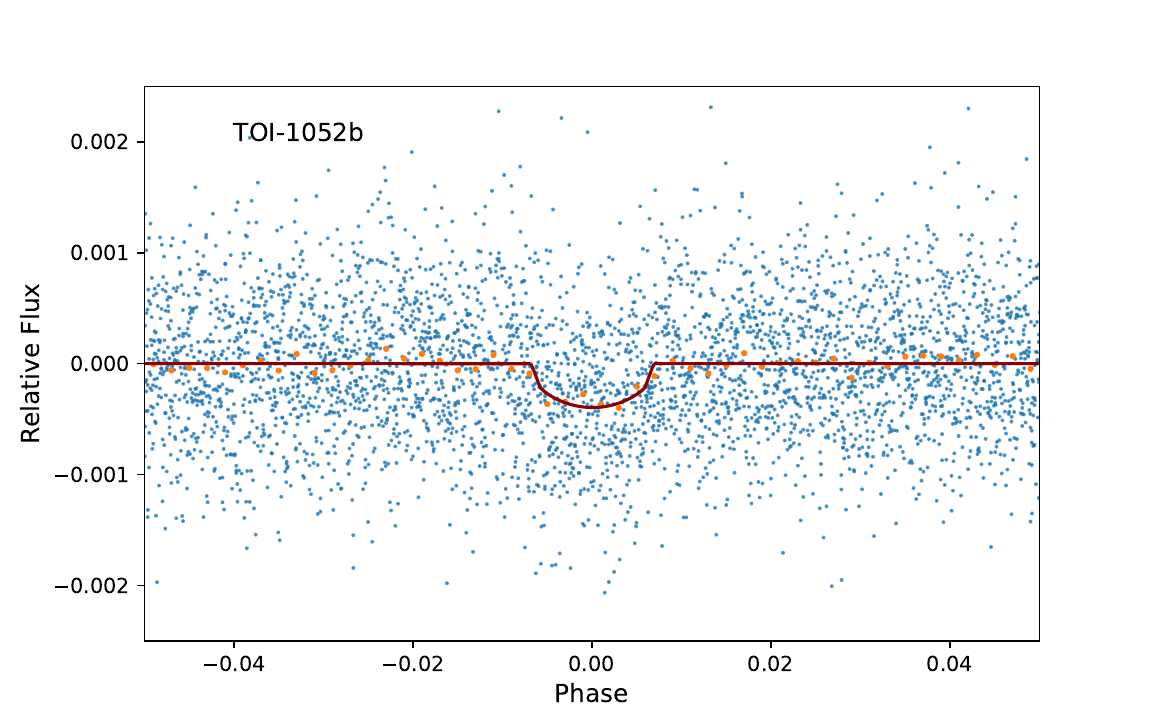}
    \caption{Fig. \ref{fig:lcall} data phase-folded on the best fitting period for TOI-1052~b with the best fit model overplotted. Binned datapoints of width 0.002 in phase are shown.  \label{fig:lcphase}}
  \end{center}
\end{figure}

We used the publicly available Presearch Data Conditioning Simple Aperture Photometry \citep[PDC-SAP,]{Twicken-10,Stumpe2012,Stumpe2014,Smith-12} light curves provided by the SPOC for the transit  modeling. Fig. \ref{fig:lcall} shows the 2 min cadence \textit{TESS} light curve, and Fig. \ref{fig:lcphase} shows the phase-folded curve for TOI\,1052b. We also identified the ~9.14 day transit event independently using the Transit Least Squares (TLS) algorithm \cite{2019A&A...623A..39H} with a Signal detection efficiency (SDE) of 19.67. No further significant periodic signal was detected in the lightcurve.

\begin{figure}
  \begin{center}
    \includegraphics[scale=0.50]{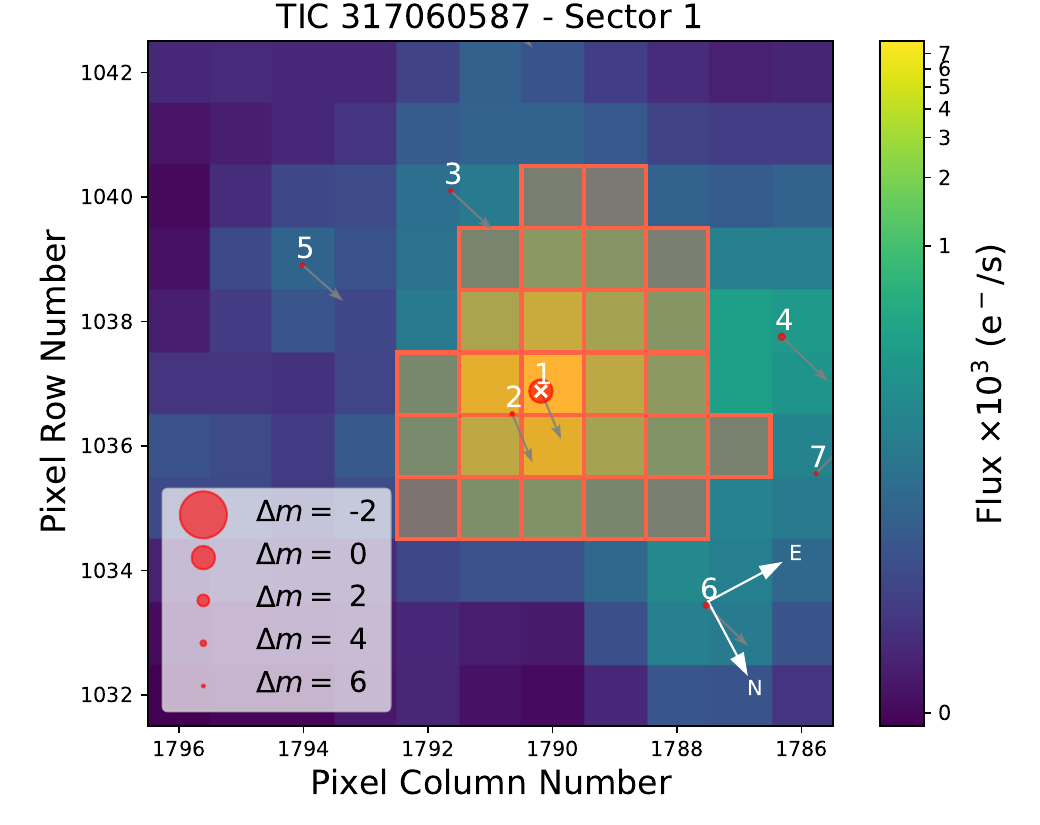}
    \includegraphics[scale=0.50]{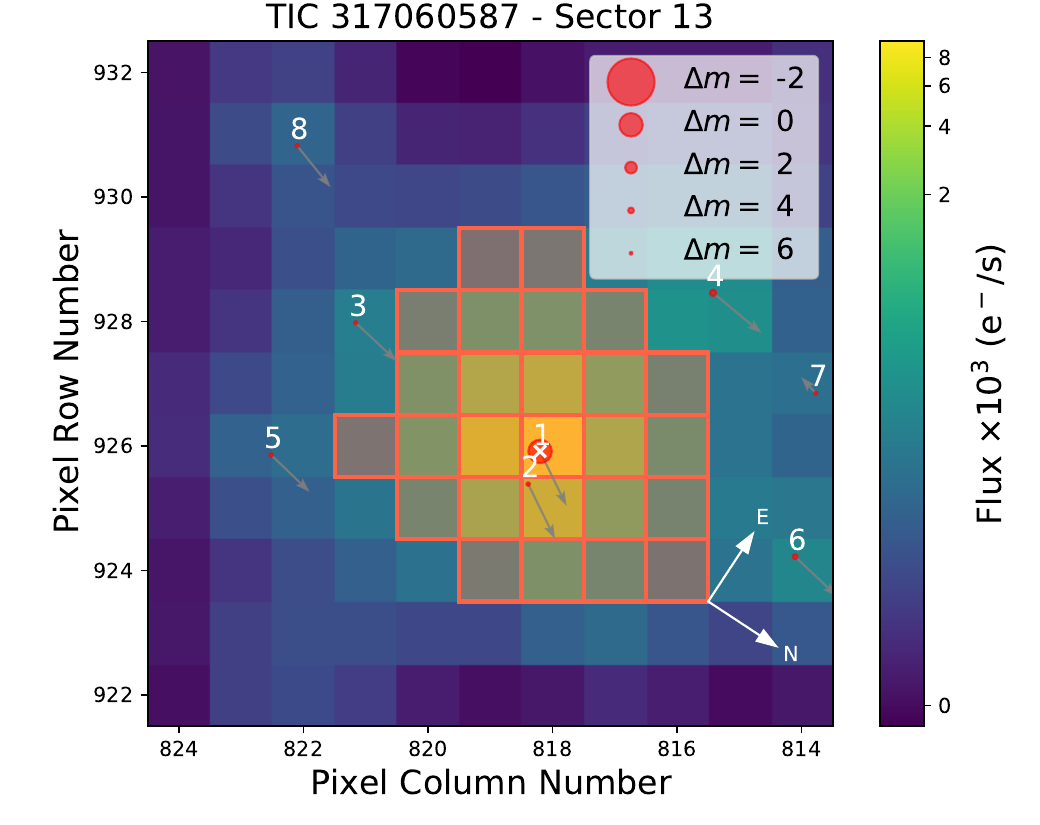}
    \caption{\textit{TESS} pixel data with GAIA DR3 data sources overplotted in sector 1 (top) and 13 (bottom). TOI-1052 is marked with a white cross and the magnitude contrast is shown as red circles. Arrows show the proper motion of each star. Aperture pixels are highlighted in red. Star 2 is a potential bound companion to TOI-1052 with consistent distance and proper motion. \label{tpftess}}
  \end{center}
\end{figure}

Following \cite{2020A&A...635A.128A} we searched for sources of flux contamination by over-plotting the GAIA DR3 catalogue to the TESS Target Pixel Files (TPF), shown in Fig. \ref{tpftess}. According to GAIA DR3 \citep{GaiaDR32022}, one star exists inside the TESS aperture in addition to TOI-1052 (Gaia DR3 6357524189130821376), with a magnitude contrast in the Gaia pass-band of 5.38, leading to negligible dilution of the transit. The nearby star is separated from TOI-1052 by 11.51”, is at a distance of 128.9~pc consistent with the distance to TOI-1052, and has similar proper motion to TOI-1052 as shown by the proper motion vectors in Fig. \ref{tpftess}. As such, the stars potentially form a bound system with a projected sky separation of \mytilde1500~AU. The secondary star has a temperature of 3600~K derived from the Gaia passbands \citep{Gaiastellar2022}.

\subsection{High-resolution imaging}

 \begin{figure}
  \begin{center}
    \includegraphics[scale=0.55]{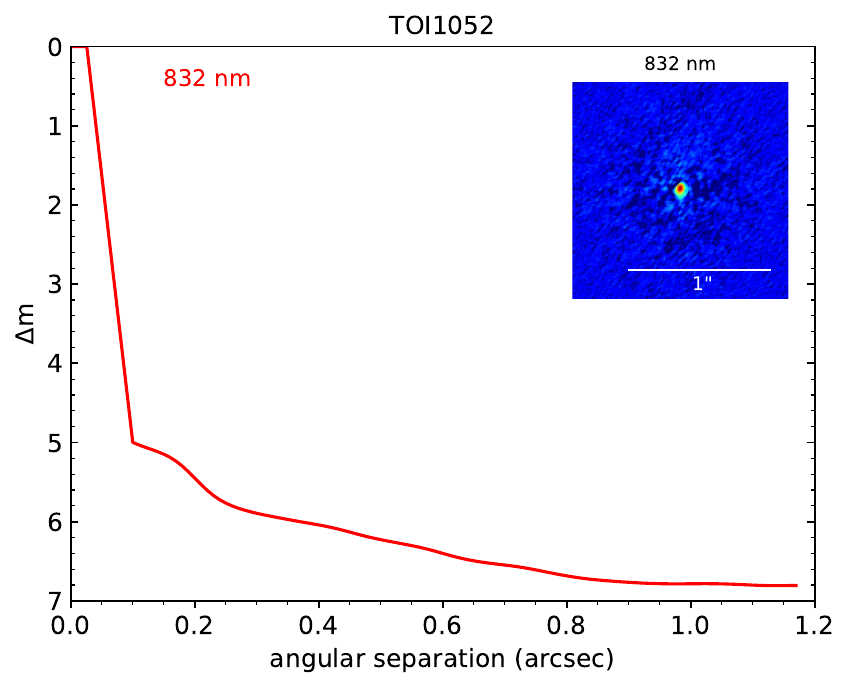}
    \caption{5$\sigma$ contrast curve for high resolution imaging observations with Zorro/Gemini. The 832 nm reconstructed image is shown in the upper right.  \label{himage} }
 
  \end{center}
\end{figure}

TOI-1052 was observed on 2021 July 07 UT using the Zorro speckle instrument on the Gemini South 8-m telescope \citep{Scott2021, Howell2022}.  Zorro provides simultaneous speckle imaging in two bands (562 nm and 832 nm) with output data products including a reconstructed image with robust contrast limits on companion detections. Five sets of 1000 $\times$ 0.06\,s images were obtained at 832 nm only and processed in the standard reduction pipeline \citep[see ][]{Howell2011}. TOI-1052 was found to have no close companions within the angular and 5$\sigma$ contrast limits (5-7 magnitudes below the target star) achieved by the observations (see Fig. \ref{himage}).  The angular limits from the 8-m Gemini telescope range from the diffraction limit (20 mas) out to 1.2”. At the distance of TOI-1052 (d=129.8 pc) these angular limits correspond to spatial limits of 2.6 to 155.8 AU.

\subsection{HARPS follow-up}

We collected 53 HARPS high-resolution spectra of TOI-1052 in three observation programs, between 2021-05-24 and 2021-09-22. The spectrograph is mounted at the ESO 3.6m telescope at La Silla Observatory, Chile \citep[][]{Mayor-13} and optimised to measure high precision RVs. The observations were carried out as part of the NCORES large program (46 obs, ID 1102.C-0249, PI: Armstrong), with supplementary observations from the NGTS-HARPS Program (5 obs, ID 0105.C-0773(A), PI: Wheatley) and the Small planets inside and out  program (2 obs, ID: 1106.C-0597(A), PI: Gandolfi).

We used the HARPS Data Reduction Software (DRS) to reduce the data, using a G0 template in order to measure RVs using a cross-correlation function (CCF) \cite[][]{Pepe-02,Baranne-96}. The spectrum signal-to-noise ratio (SNR) is approx. 40 per pixel leading to a mean photon-noise uncertainty of 2.06 m s$^{-1}$.  The DRS was used to measure the full width half maximum (FWHM), the line bisector, and the contrast of the CCF, as well as several activity indicators. The mean $log(R' _{HK})$ of the star is $-5.32\pm0.02$ implying a relatively low magnetic activity level. 

\begin{figure*}
  \begin{center}
    \includegraphics[width=\textwidth]{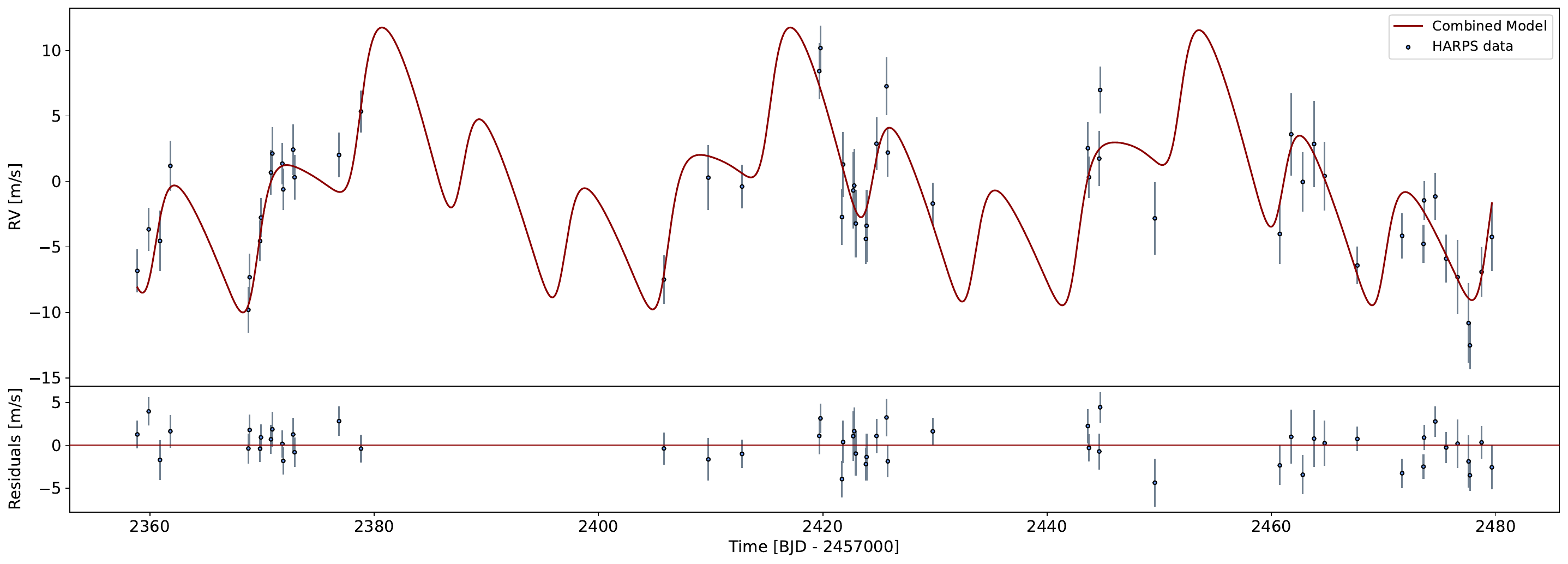}
    \caption{The full radial velocity HARPS timeseries showing the combined best fit model from planets b and c in red. Residuals after subtracting the model are shown below. \label{fig:rvall}}
  \end{center}
\end{figure*}

The full RV timeseries is shown in Fig. \ref{fig:rvall}.

\section{Host star fundamental parameters}
\label{sec:star}

\subsection{Spectroscopic Analysis}
The stellar spectroscopic parameters ($T_{\mathrm{eff}}$, $\log g$, microturbulence, [Fe/H]) were estimated using the ARES+MOOG methodology. The methodology is described in detail in \citet[][]{Sousa-21, Sousa-14, Santos-13}. To consistently measure the equivalent widths (EW) we used the latest version of ARES \footnote{The latest version, ARES v2, can be downloaded at https://github.com/sousasag/ARES} \citep{Sousa-07, Sousa-15}. The list of iron lines is the same as the one presented in \citet[][]{Sousa-08}. For this we used a co-added HARPS spectrum of TOI-1052. In this analysis we use a minimization process to find the ionization and excitation equilibrium to converge on the best set of spectroscopic parameters. This process makes use of a grid of Kurucz model atmospheres \citep{Kurucz-93} and the radiative transfer code MOOG \citep{Sneden-73}. We also derived a more accurate trigonometric surface gravity using recent GAIA data following the same procedure as described in \citet[][]{Sousa-21} which provided a consistent value when compared with the spectroscopic surface gravity. The resulting spectroscopic parameters are given in Table \ref{tab:stellar}. The derived temperature of $6146\pm62$K is indicative of a late F star as opposed to the G0 type specified in the literature \citep{1975mcts.book.....H}, but not different at high enough confidence for us to reclassify the spectral type.

The abundances of the following elements were also derived using the same tools and models as for the stellar spectroscopic parameters: Mg, Al, Ti, Si, and Ni \citep[detailed in e.g.][]{Adibekyan-12, Adibekyan-15}, neutron capture elements (used later to obtain ages) as explained in \citet{DelgadoMena-17}, and C and O \citep[following][]{Bertrandelis-15,DelgadoMena-21}. Although the equivalent widths (EWs) of the spectral lines were automatically measured with ARES, we performed careful visual inspection of the EWs measurements.

\subsection{Stellar Mass, Radius and Age}

The stellar mass, radius and age were estimated from the spectroscopically derived parameters using the \texttt{PARAM} 1.3 web-interface\footnote{http://stev.oapd.inaf.it/cgi-bin/param\_1.3} \citep[][]{daSilva-06}, leading to $R_{\star}=1.264 \pm 0.033 R_\odot$, $M_{\star}=1.204 \pm 0.025M_\odot$, and Age $= 2.3 \pm 1.0$\ Gyr. As an alternative we also estimated the stellar mass from the \texttt{PARAM} 1.3 values using the calibration presented in \citet[][]{Torres-2010} which provided a consistent result ($M_{\star\mathrm{, Torres}} = 1.19 \pm 0.03  M_\odot $).

As an independent determination of the basic stellar parameters, we also performed an analysis of the broadband spectral energy distribution (SED) of the star together with the {\it Gaia\/} EDR3 parallax \citep[with no systematic offset applied; see, e.g.,][]{StassunTorres2021}, in order to determine an empirical measurement of the stellar radius, following the procedures described in \citet{Stassun:2016,Stassun:2017,Stassun:2018}. We pulled the $B_T V_T$ magnitudes from {\it Tycho-2}, the $JHK_S$ magnitudes from {\it 2MASS}, the W1--W4 magnitudes from {\it WISE}, and the $G G_{\rm BP} G_{\rm RP}$ magnitudes from {\it Gaia}. Together, the available photometry spans the stellar SED over the wavelength range 0.4--22~$\mu$m.

We performed a fit using Kurucz stellar atmosphere models, with the main parameters being the effective temperature ($T_{\rm eff}$), surface gravity ($\log g$), and metallicity ([Fe/H]), for which we adopted the spectroscopically determined values. The remaining free parameter is the extinction $A_V$, which we limited to the maximum line-of-sight value from the Galactic dust maps of \citet{Schlegel:1998}. The resulting fit has a reduced $\chi^2$ of 1.0 and best fit $A_V = 0.04 \pm 0.04$. Integrating the model SED gives the bolometric flux at Earth, $F_{\rm bol} = 4.123 \pm 0.096 \times 10^{-9}$ erg~s$^{-1}$~cm$^{-2}$. Taking the $F_{\rm bol}$ and $T_{\rm eff}$ together with the {\it Gaia\/} parallax, gives the stellar radius, $R_\star = 1.293 \pm 0.030$~R$_\odot$. In addition, we can again estimate the stellar mass from the empirical relations of \citet{Torres:2010}, giving $M_\star = 1.20 \pm 0.07$~M$_\odot$.

All of our methods of stellar parameter estimation produce consistent results. We adopt the \texttt{PARAM} 1.3 values going forwards, and these are listed in Table \ref{tab:stellar}.

\subsection{Rotational period and age}
\label{sect:rot}
Our \texttt{PARAM} fit led to an estimated isochrone age of $= 2.3 \pm 1.0$\ Gyr. We are also able to estimate the stellar age via the chemical clocks method \citep[see][]{DelgadoMena2019}. The ages estimated from different chemical clocks (together with the Teff and [Fe/H]) are listed in Table \ref{tab:clocks}, giving a weighted average of $1.9\pm0.3$ Gyr consistent with the \texttt{PARAM} age. This small error bar just reflects the good agreement between the ages obtained from different chemical clocks. We conservatively adopt the \texttt{PARAM} age with its larger error bar, given the uncertainties associated with stellar age estimation.

We do not find any evidence of periodic variability indicative of rotation in the \textit{TESS} lightcurves.

Through the FWHM of the HARPS spectra CCF we are able to estimate the projected rotation velocity $v\sin i$ for the star. The mean FWHM across the spectra is 9.12~kms$^{-1}$. Using a calibration similar to the one presented in \citet[][and references therein]{2002A&A...392..215S,maldonado2017hades,2019A&A...629A..80H} this FWHM implies a $v\sin i$ of $5.63\pm0.5$~kms$^{-1}$. We also re-derived the $v\sin i$ by performing spectral synthesis with MOOG on 36 isolated iron lines and by fixing all the stellar parameters, macroturbulent velocity, and limb-darkening coefficient \citep{CostaSilva-20}, leading to a consistent value of v$\sin i$ = 5.0$\pm$0.9~km/s, which we adopt. The  linear limb-darkening coefficient (0.7) was determined using the ExoCTK package \citep{matthew_bourque_2021} using the determined stellar parameters. The macroturbulent velocity (4.4 km/s) was determined using the temperature and gravity dependent empirical formula from \citet{Doyle2014}.

 We estimated the (projected) rotation period directly via the spectroscopic $v\sin i$ and the $R_\star$ determined above, which gives $P_{\rm rot} / \sin i = 12.8 \pm 2.3$~d. Assuming the stellar orbital inclination is $i \approx 90^\circ$, then this would represent approximately the true rotation period.

\begin{table}
 \centering
\caption{\label{tab:clocks} Chemical clock age estimates \citep[see][Table 10]{DelgadoMena2019}.}
	\begin{tabular}{lr}
	\hline\hline
	\noalign{\smallskip}
	Clock	&	Value (Gyr)	\\
	\hline
$\mathrm{[Y/Zn]}$     &     $2.1\pm0.5$ \\
$\mathrm{[Y/Ti]}$    &     $1.6\pm0.7$\\
$\mathrm{[Y/Mg]}$     &     $1.4\pm0.6$\\
$\mathrm{[Sr/Ti]}$    &     $2.3\pm1.3$\\
$\mathrm{[Sr/Mg]}$    &     $2.0\pm1.1$\\
$\mathrm{[Y/Si]}$     &     $1.6\pm0.7$\\
$\mathrm{[Sr/Si]}$    &     $1.8\pm1.2$\\
$\mathrm{[Y/Al]}$    &    $2.7\pm0.3$\\
Weighted Mean   &    $1.9\pm0.3$  \\  
 	\noalign{\smallskip}
	\hline
 	\noalign{\smallskip}
    \end{tabular}
    
\end{table}

\subsection{Signal identification}
\label{sec:signals}

\begin{figure}
  \begin{center}
    \includegraphics[width=0.5\textwidth]{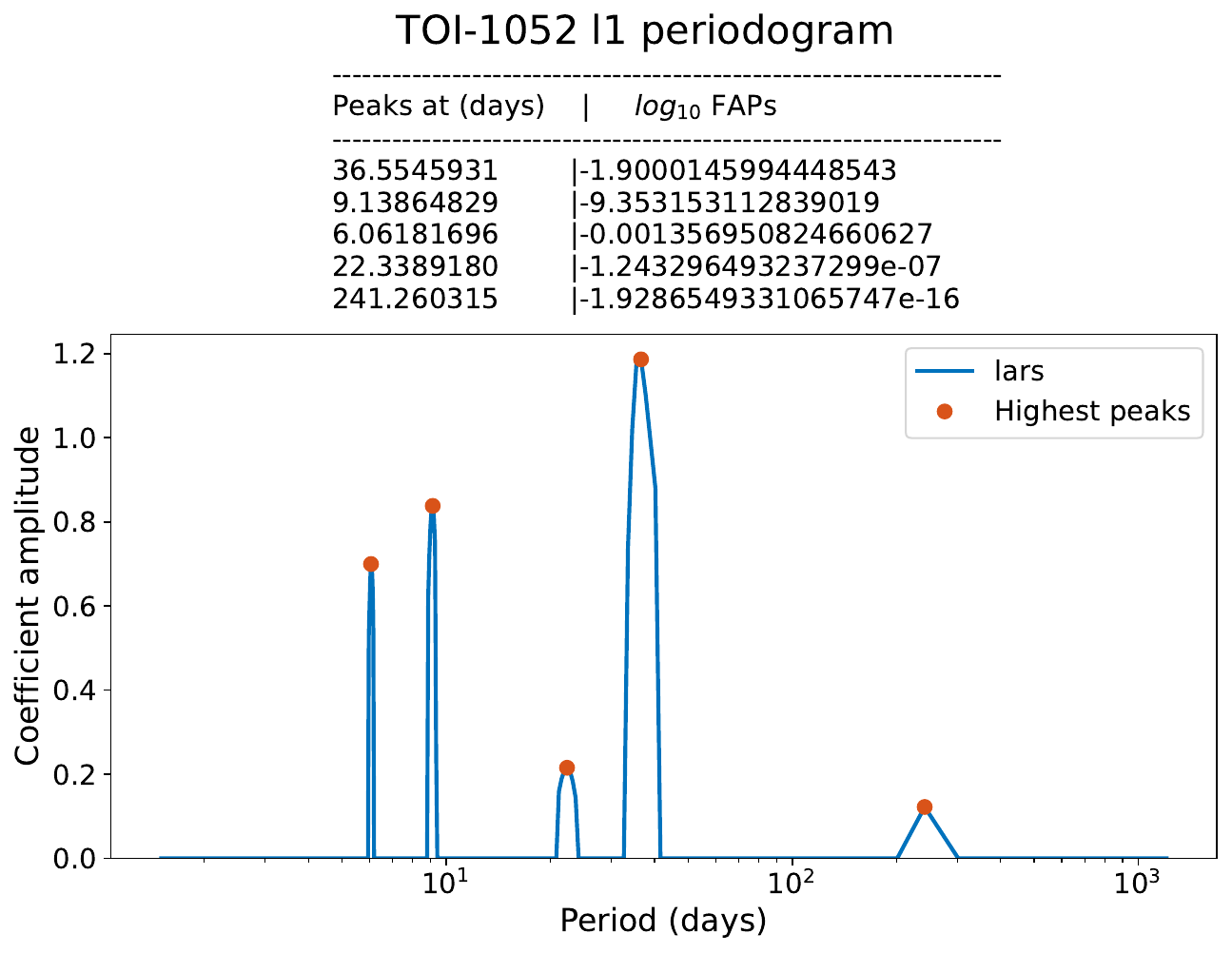}
    \caption{ $l_1$ Periodogram as discussed in Section \ref{sec:signals} showing significant peaks at 9.14 and 36.6d. Peak values and false alarm probabilities (FAPs) are shown above the periodogram. \label{fig:l1}}
  \end{center}
\end{figure}

\begin{figure}
  \begin{center}
    \includegraphics[scale=0.45]{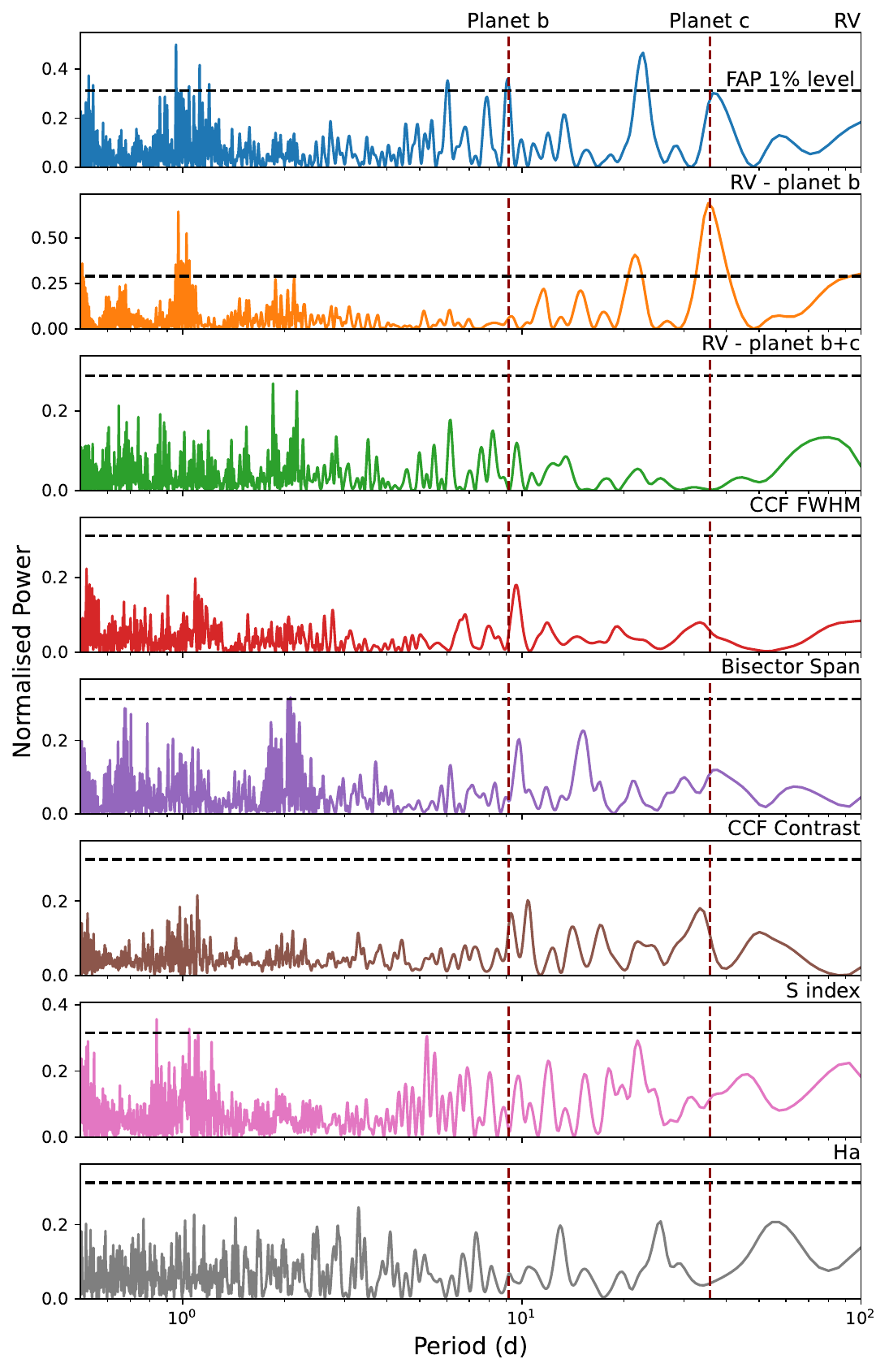}
    \caption{Lomb-Scargle periodogram of HARPS RVs and activity indicators. The orbital period of each planet is shown as vertical dashed lines. Horizontal dashed lines show the FAP 1\% level. A 22d artefact is seen in the raw RVs at the top, but vanishes when both planet models are removed.} \label{fig:periodogram} 
  \end{center}
\end{figure}

We computed the $l_1$ periodogram \citep[e.g.,][]{Hara-17,Hara-20} to find periodicities in the RV data. The $l_1$  periodogram uses the theory of compressed sensing adapted for handling correlated noise to analyze the radial velocity without the estimation of the frequency iteratively, see \citet{Hara-17,Hara-20} for more information. A fundamental difference of the $l_1$ periodogram over the typically used Lomb-Scargle \citep{Lomb1976,Scargle1982,Vanderplas2018} is that all possible frequencies are tested simultaneously. This method reduces aliases in the periodogram. Fig.~\ref{fig:l1} shows two significant signals, considering the model noise with a 1.5 m/s jitter noise, consistent with the eventual jitter found by our best fit model in Table \ref{table:posterior}. Both the 9.14 day transiting planet period and an additional 36.6 day period were found to be significant with a False Alarm Probability (FAP) < 1.0 $\% $ for the 9.14 day and 1.2\% for the 36.6 day, as opposed to the other shown peaks which have FAP > 99\%. Lomb-Scargle periodograms of the RVs with and without the planet signals, FWHM, bisector span (BIS), log RHK and CCF contrast, calculated with \texttt{astropy} \citep{Astropy3_2022}, are shown in Fig. \ref{fig:periodogram}, to investigate the planet peaks further and consider a potential activity source for the significant signals. No significant power is found at the 9 day or 36 day periods in any of the indicators. Fig. \ref{fig:periodogram} also demonstrates that two periodic components are required to model the RVs, with no further periodic signals found once the two planets are removed. Note the the initially most significant peak seen in the RVs in the Lomb-Scargle periodogram is at 22d, which is seen with low significance in the $l_1$ periodogram. The 22d peak is an artefact arising from both the 9.14 day and 36 day planet peaks and vanishes when both planets are removed. Similarly, the 6d signal seen in both periodograms is an artefact of the planet b peak.

Given the robust detection of the transiting planet in the radial velocities, we are able to confirm the known planetary candidate as TOI-1052~b. Given that there is no sign of stellar activity in any indicator at the 36.6 day additional period, and this period does not match the estimated stellar rotation period or its harmonics, we claim this as an additional planet in the system, TOI-1052~c. The joint fit of photometry and spectroscopy in Section \ref{sect:model} finds a period for planet c of $35.81^{+0.45}_{-0.38}$d. TOI-1052~c is just within the 4:1 resonance of planet b. The system dynamics are discussed in Section \ref{sect:dyn}.

\begin{table}
    \small
    \caption{Prior distributions used in our joint fit model, fully described in Section \ref{sect:model}. The priors are created using distributions in \texttt{PyMC3} with the relevant inputs to each distribution described in the table footer. Fit results and derived parameters can be found in Table \ref{table:posterior}}
	\label{tab:jointfit}
	\begin{threeparttable}
	\begin{tabular}{l p{1.3cm} l}
	\toprule
	\textbf{Parameter} & \textbf{(unit)} & \textbf{Prior Distribution}  \\
	\midrule
	\multicolumn{3}{l}{\textbf{Planet b}} \\
	Period $P_b$                      & (days)               & $\mathcal{N}(9.13966, 0.001)$       \\
	Ephemeris $t_{0,b}$                 & (BJD-2457000)        & $\mathcal{N}(1332.9448, 0.02)$   \\
	Radius $\log{(R_b)}$                   & (log \mbox{R$_{\odot}$}) & $\mathcal{N}(-3.733^*, 1.0)$       \\
        Impact Parameter $b_b$         &                    &       $\mathcal{U}(0, 1+R_b/R_{*})$         \\
         $e_b\sin\omega_b$                &                    & $\mathcal{U}(\textrm{Unit disk})$                             \\
	$e_b\cos\omega_b$                   &         & $\mathcal{U}(\textrm{Unit disk})$                            \\
	$K_b$                     & (m\,s$^{-1}$)        & $\mathcal{U}(0.0, 50.0)$              \\
	\midrule
	\multicolumn{3}{l}{\textbf{Planet c}} \\
	Period $P_c$                      & (days)               & $\mathcal{N}(35.97306, 4.0)$        \\
	Ephemeris $t_{0,c}$                 & (BJD-2457000)        & $\mathcal{N}(2423.3168, 20.0)$   \\
	$e_c\sin\omega_c$                &                    & $\mathcal{U}(\textrm{Unit disk})$                             \\
	$e_c\cos\omega_c$                   &         & $\mathcal{U}(\textrm{Unit disk})$                          \\
	$K_c$                     & (m\,s$^{-1}$)        & $\mathcal{U}(0.0, 50.0)$              \\
	\midrule
	\multicolumn{3}{l}{\textbf{Star}} \\
	Mass $M_{*}$                    & (\mbox{M$_{\odot}$}) & $\mathcal{N_B}(1.204, 0.025, 0.0, 3.0)$  \\
	Radius $R_{*}$                  & (\mbox{R$_{\odot}$}) & $\mathcal{N_B}(1.264, 0.033, 0.0, 3.0)$   \\
	\midrule	
	\multicolumn{3}{l}{\textbf{Photometry}} \\
	TESS mean              &                    & $\mathcal{N}(0.0, 1.0)$                \\
	$\log{(\rm{Jitter)}}$          &       (\mbox{m\,s$^{-1}$})    & $\mathcal{N}(-7.40^{\dagger}, 10)$                \\
        \midrule
	\multicolumn{3}{l}{\textbf{HARPS RVs}} \\
	Offset                          & (m\,s$^{-1}$)        & $\mathcal{N}(54945.0, 10.0)$        \\
	$\log{(\rm{Jitter)}}$           & (m\,s$^{-1}$)        & $\mathcal{N}(0.37^{\dagger}, 5.0)$  \\
	\bottomrule
	\end{tabular}
	\begin{tablenotes}
	\item \textbf{Distributions:}
	\item $\mathcal{N}(\mu, \sigma)$: a normal distribution with a mean $\mu$ and a standard deviation $\sigma$;
	\item $\mathcal{N_B}(\mu, \sigma, a, b)$: a bounded normal distribution with a mean $\mu$, a standard deviation $\sigma$, a lower bound $a$, and an upper bound $b$ (bounds optional);
	\item $\mathcal{U}(a, b)$: a uniform distribution with a lower bound $a$, and an upper bound $b$.
	\item \textbf{Prior values:}
	\item $^*$ equivalent to $0.5(\log{(D)}) + \log{(R_{*})}$ where $D$ is the transit depth (ppm multiplied by $10^{-6}$) and $R_{*}$ is the mean of the prior on the stellar radius (\mbox{R$_{\odot}$});
	\item $^{\dagger}$ equivalent to the log of the minimum error on the HARPS data (m\,s$^{-1}$), or the mean error on the \textit{TESS} data. We fit a log value to enforce an broad, non-zero prior covering several orders of magnitude.
	\end{tablenotes}
	\end{threeparttable}
\end{table}

\begin{table*}
\begin{minipage}{12cm}
\caption{TOI-1052 fit and derived parameters: Median and 68\% confidence interval.}
\label{table:posterior}
\begin{tabular}{lllllr}\hline

Model &  & \multicolumn{2}{l}{With eccentricity (adopted)} & \multicolumn{2}{l}{Fixed eccentricity} \\

\smallskip\\\multicolumn{2}{l}{\underline{System Parameters:}}&\smallskip\\

~~~~$u_{1,TESS}$\dotfill & &\multicolumn{2}{l}{$0.80^{+0.61}_{-0.55}$}& \multicolumn{2}{l}{$0.85^{+0.59}_{-0.57}$}\\
~~~~$u_{2,TESS}$\dotfill & &\multicolumn{2}{l}{$-0.07^{+0.52}_{-0.49}$}& \multicolumn{2}{l}{$-0.12^{+0.54}_{-0.48}$}\\
~~~~\textit{TESS}$_\textrm{offset}$\dotfill & ppm &\multicolumn{2}{l}{$-7.9^{+3.3}_{-3.3}$} & \multicolumn{2}{l}{$-7.9^{+3.2}_{-3.2}$}\\
~~~~$\sigma _{TESS}$\dotfill &ppm &\multicolumn{2}{l}{$61.0^{+2.3}_{-2.3}$} & \multicolumn{2}{l}{$61.0^{+2.3}_{-2.3}$}\\
~~~~$\sigma _{HARPS}$\dotfill &m/s& \multicolumn{2}{l}{$0.79^{+0.61}_{-0.65}$} & \multicolumn{2}{l}{$1.62^{+0.42}_{-0.42}$}\\
~~~~Systemic RV\dotfill & m/s & \multicolumn{2}{l}{$54946.02^{+0.38}_{-0.39}$} & \multicolumn{2}{l}{$54945.75^{+0.38}_{-0.39}$} \\

\smallskip\\\multicolumn{2}{l}{\underline{Planetary Parameters:}}&b&c&b&c\smallskip\\

~~~~$P$ \dotfill &days &$9.139703^{+0.000190}_{-0.000197}$&$35.806^{+0.453}_{-0.381}$ & $9.139664^{+0.000191}_{-0.000209}$&$35.779^{+0.466}_{-0.420}$\\
~~~~$T_0$ \dotfill &$\textrm{BJD}-2457000$  &$1332.9442^{+0.0057}_{-0.0055}$& $2425.13^{+1.63}_{-1.64}$ & $1332.9454^{+0.0060}_{-0.0056}$& $2427.46^{+0.54}_{-0.50}$\\
~~~~$K$\dotfill & m/s &$4.70^{+0.46}_{-0.46}$&$6.11^{+0.80}_{-0.69}$ & $4.45^{+0.54}_{-0.52}$&$5.33^{+0.62}_{-0.63}$\\
~~~~$e$\dotfill & &$0.180^{+0.090}_{-0.071}$& $0.237^{+0.090}_{-0.082}$ & 0 (fixed) & 0 (fixed) \\
~~~~$\omega$\dotfill & rad &$-2.07^{+0.24}_{-0.42}$& $-0.79^{+0.47}_{-0.53}$ & - & -  \\
~~~~$a/R_{\star}$  &  &$15.51^{+0.43}_{-0.41}$& $38.6\pm1.1$ &$15.53^{+0.43}_{-0.41}$& $38.6\pm1.1$ \\
~~~~$i$\dotfill & deg &$87.53^{+0.24}_{-0.20}$& - &$87.55^{+0.46}_{-0.29}$& - \\
~~~~$R_p/R_{\star}$\dotfill & &$0.0194^{+0.0018}_{-0.0015}$& - & $0.0184^{+0.0013}_{-0.0014}$& -  \\
~~~~$b$\dotfill & &$0.808^{+0.052}_{-0.079}$& - & $0.727^{+0.052}_{-0.085}$& - \\

\smallskip\\\multicolumn{2}{l}{\underline{Derived parameters:}}&b&c&b&c\smallskip\\

~~~~$M_p$\dotfill & $M_{\oplus}$   &$16.9^{+1.7}_{-1.7}$& - &$16.4^{+2.0}_{-1.9}$& - \\
~~~~$R_p$\dotfill & $R_{\oplus}$&$2.87^{+0.29}_{-0.24}$& - & $2.72^{+0.22}_{-0.22}$& - \\
~~~~$\rho _p$\dotfill & $g/cm^3$   &$3.93^{+1.7}_{-1.3}$& -  &$4.48^{+2.0}_{-1.4}$& - \\
 ~~~~$a_p$\dotfill & $AU$  &$0.09103^{+0.00062}_{-0.00063}$& $0.2263^{+0.0024}_{-0.0023}$ &$0.09102\pm0.00062$& $0.2261\pm0.0024$ \\
~~~~$S$\dotfill & $S_{\oplus}$   &$246\pm13$& $39.7^{+2.3}_{-2.2}$ &$245^{+14}_{-13}$& $39.7^{+2.3}_{-2.2}$ \\
~~~~$T_{eq}\dagger$ \dotfill & $K$   & $1135$ & $719$ & $1134$ & $719$ \\
~~~~$M_P\sin i$\dotfill &$M_{\oplus}$ & - & $34.3^{+4.1}_{-3.7}$ & - & $30.9^{+3.6}_{-3.7}$ \\
\\

\hline
\end{tabular}
   \vspace{1ex}

     {\raggedright * For TOI 1052c $T_0$ corresponds to the time when the planet would have transited. \\
     $\dagger \textrm{Assuming a Bond Albedo of 0.3.}$}
\end{minipage}
\end{table*}

\section{Joint modelling}
\label{sect:model}

\begin{figure}
  \begin{center}
    \includegraphics[scale=0.38]{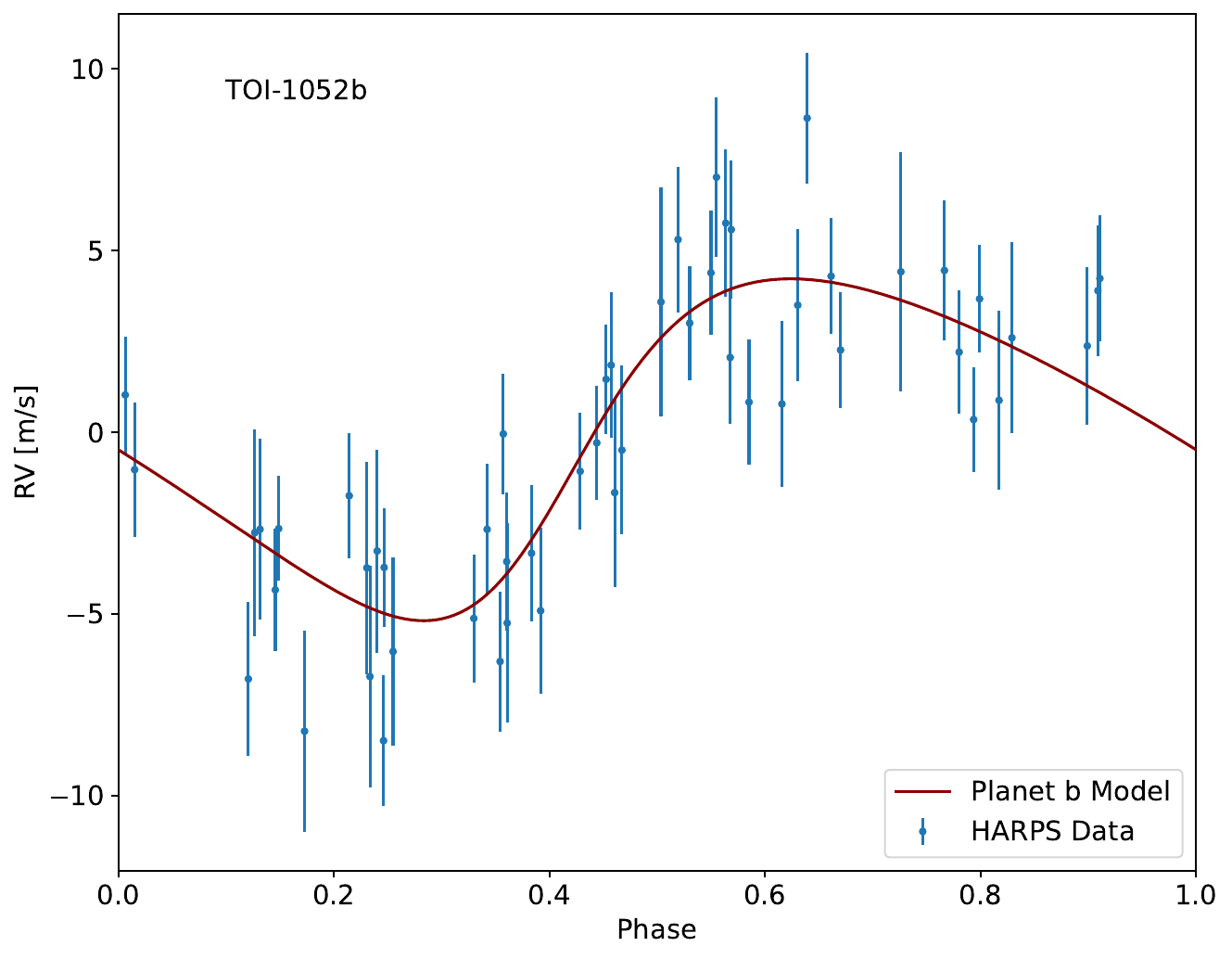}
    \caption{Radial velocities phase-folded at the best fitting period of TOI-1052~b, with best fit model overplotted in red.  \label{fig:rvphaseb}}
  \end{center}
\end{figure}

\begin{figure}
  \begin{center}
    \includegraphics[scale=0.38]{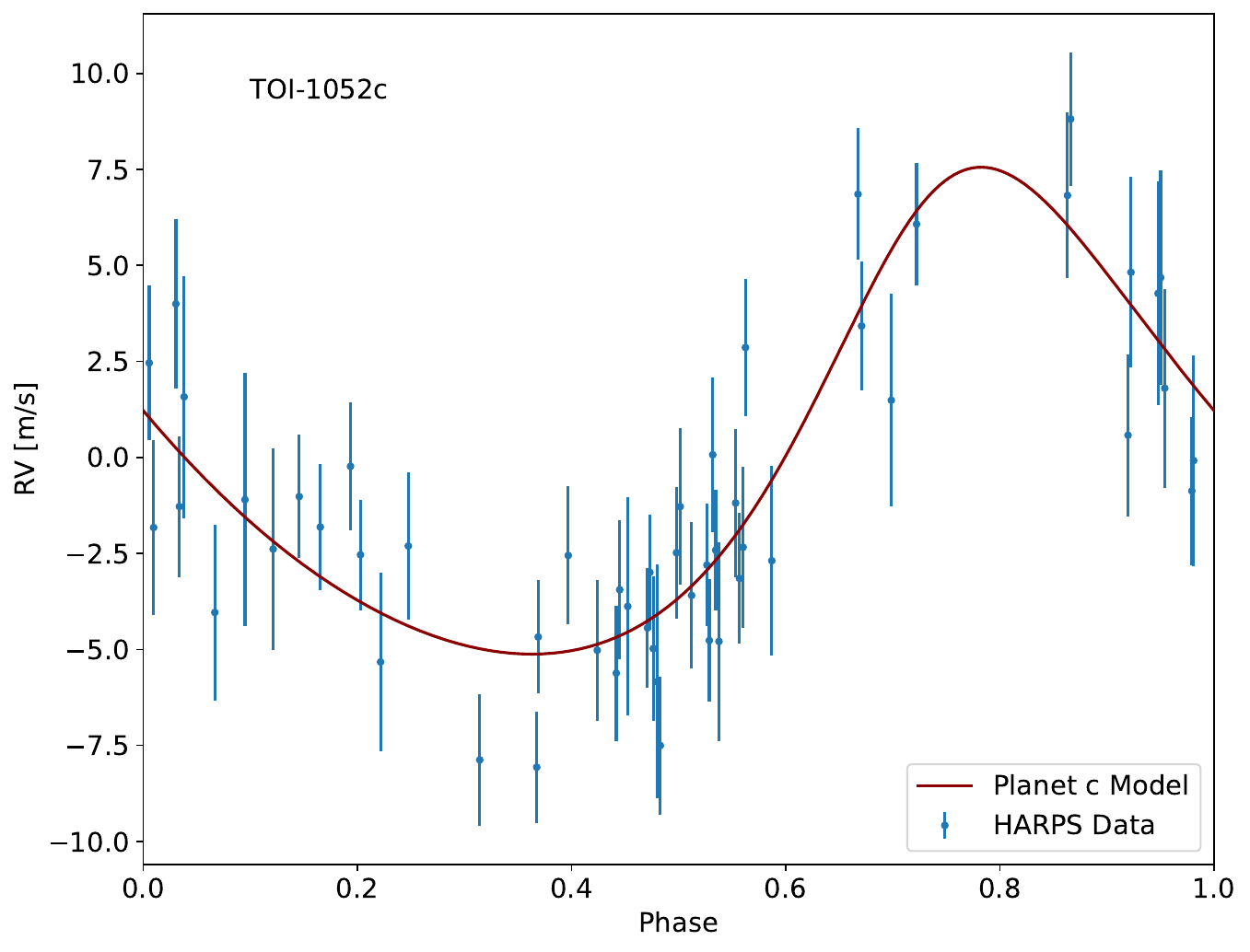}
    \caption{Radial velocities phase-folded at the best fitting period of TOI-1052~c, with best fit model overplotted in red. \label{fig:rvphasec}}
  \end{center}
\end{figure}

The photometry from {\it TESS} and spectroscopy from HARPS were combined in a joint fit using the \texttt{exoplanet} \citep{exoplanet:joss} code framework. This package also makes use of \texttt{starry} \citep{starryLuger2019} and \texttt{PYMC3} \citep{SalvatierPymc3}. The photometric model is adjusted to account for the \textit{TESS} exposure time of 2 minutes.

The model constructs two Keplerian orbits, one for each planet, with orbital period $P$, epoch $t_0$, impact parameter $b$, eccentricity $e$ and angle of periastron $\omega$ as free parameters determining the orbit. The orbital period and epoch are drawn from Gaussian prior distributions with a mean drawn from initial fits and a standard deviation of 0.001 and 4 days for planets b and c respectively, approximately 10 times larger than the eventual errors on those parameters. The impact parameter is drawn uniformly between $0$ and $1+R_p/R_\star$, where $R_p$ is the planetary radius. $e$ and $\omega$ are drawn via sampling $e\sin\omega$ and $e\cos\omega$ from a unit disk distribution then deriving $e$ and $\omega$. Additionally the stellar mass and radius are allowed to vary in a Gaussian distribution according to their values from Section \ref{sec:star}.

Once the orbit is defined, the planet to star radius ratio $R_p/R_\star$ and radial velocity semi-amplitude $K$ are drawn from wide uniform distributions. Limb darkening parameters are drawn from the quadratic limb darkening parameterisation of \citet{kipping2013}. We introduce a systematic radial velocity offset, a \textit{TESS} photometry offset, and instrumental jitter parameters for both instruments as extra parameters. Jitter is drawn from a broad Gaussian distribution in log-space to allow for a wide range of orders of magnitude, and is then added to the measured instrumental noise in quadrature.

We do not include a model for the stellar noise, apart from the jitter term in the RVs, as no significant periodic signal was found in either the RVs, stellar activity indicators or photometry aside from the two planetary signals.

We use a No U-Turn Sampler (NUTS) variant of the Hamiltonian Monte Carlo (HMC) algorithm to draw samples from the posterior chain, for which we use 12 chains each with 5,000 steps for a total of 60000 iterations. We treat the first 1500 samples drawn from each chain as burn-in and subsequently discard them. The resulting Gelman-rubin statistics \citep{GRstat1998} for each variable are $<<1.05$, demonstrating the chains have converged.  

Our initial fits revealed a marginally significant eccentricity for both planets (at $2.5\sigma$ for planet b and $2.9\sigma$ for planet c). We present fit posterior values with eccentricity, and with both planets fixed at zero eccentricity, in Table \ref{table:posterior}. The resulting planet parameters are consistent in both models. To compare the models we calculate the WAIC (widely applicable information criterion), which estimates the expected log pointwise predictive density (elpd) of the models \citep[for details on the criterion see][]{Vehtari2017,watanabe10a}. The eccentric model is slightly favoured with a higher elpd, with a difference of 4.0, although this difference is not large enough to be considered significant. We adopt the free eccentricity results going forwards.

\begin{figure}
  \centering
  \begin{tabular}{@{}cc@{}}
    \includegraphics[width=0.5\textwidth]{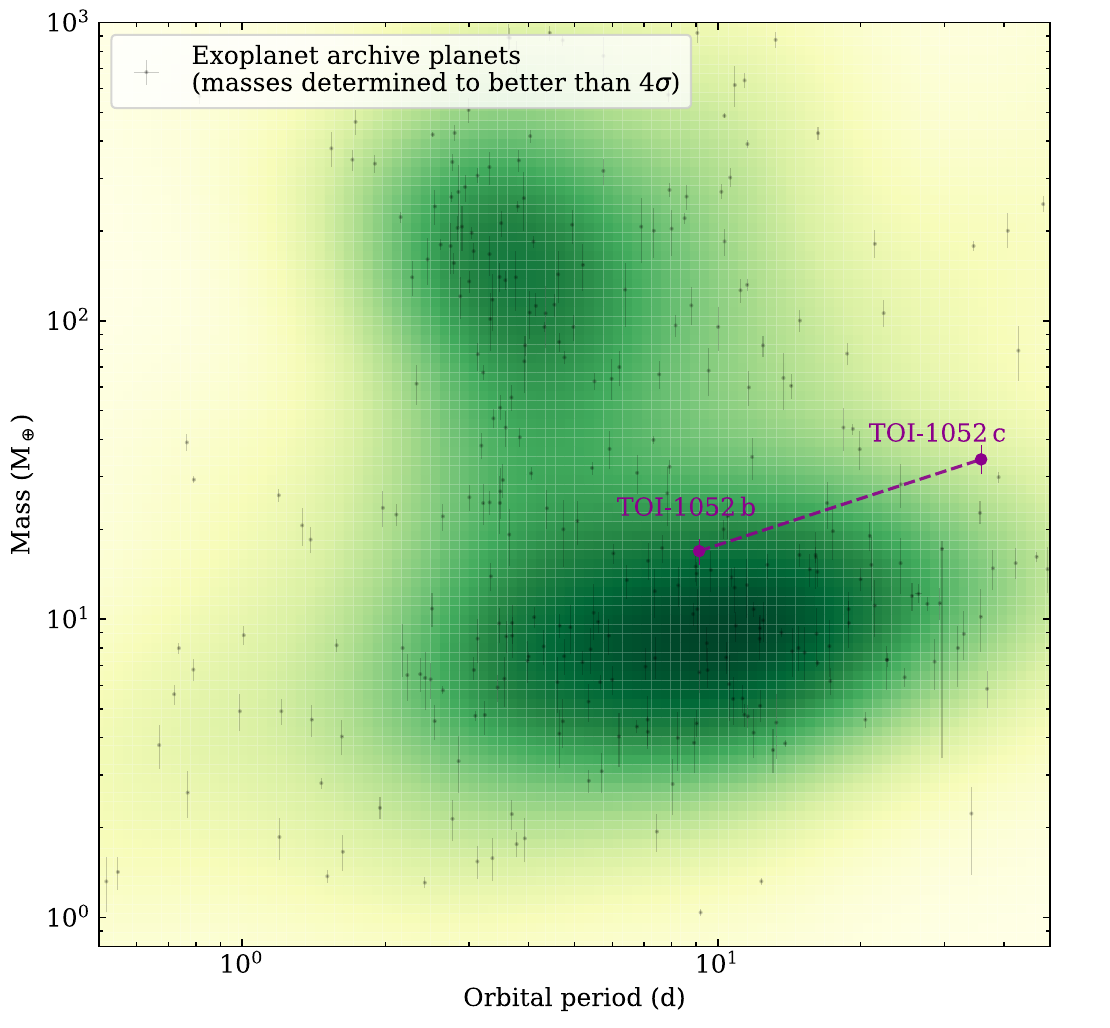}
    \end{tabular}
    \caption{Period-Mass diagram showing both planets, with TOI-1052c plotted at its $M_P\sin i$ value. The background density distribution of known exoplanets is shown in green.}
    \label{fig:p-m-relation}
\end{figure}

Through the results of this analysis, we determine that TOI-1052~b is a mini-Neptune of radius $2.87^{+0.29}_{-0.24}$ R\textsubscript{$\oplus$} and mass $16.9\pm1.7$ M\textsubscript{$\oplus$}. From these values we infer a bulk density of $3.9^{+1.7}_{-1.3}$ gcm\textsuperscript{-3}. The non-transiting planet TOI-1052~c is found to have $M\sin i_c = 34.3^{+4.1}_{-3.7}$ M\textsubscript{$\oplus$}. No evidence of transits is seen for planet c. Fig. \ref{fig:p-m-relation} shows the two planets in the context of the exoplanet population.

The near 4:1 ratio of the orbital periods, and potential eccentricity, invite questions as to whether there is a third planet in between planets b and c, forming a 1:2:4 ratio. Two planets may mimic a single planet with an eccentric orbit in radial velocity observations, although TOI-1052~c is below the $3\sigma$ significant eccentricity criterion for this issue found in \citet{2019MNRAS.484.4230W}. We could not find any evidence of such a hidden planet, which might be expected to show in the radial velocity residuals if the model is forced to be circular. Absent that evidence, we proceed with the two-planet model but note this possibility in case future observations can probe the system further.

\section{Discussion}
\label{sect:discuss}
\subsection{Dynamical analysis}
\label{sect:dyn}
 
The wide range of allowed eccentricities for both planets raise questions as to their stability and dynamical interactions. Further, the approximate $4$:$1$ ratio of the planets' orbital periods invites a more detailed analysis of a potential resonant interaction and how it affects the stability of the system.

First, because this system is a two-planet system, we can determine if it is Hill stable analytically. By using the procedure outlined in \cite{veretal2013}, which is based on the equations in \cite{donnison2006,donnison2011}, we find that the TOI-1052 is Hill stable for all planetary eccentricities in the ranges $e_{\rm b}=0.0-0.3$ and $e_{\rm c}=0.0-0.3$. In fact, the system is comfortably Hill stable: even in the scenario where $e_{\rm b} = e_{\rm c} = 0.3$, the system would be Hill stable for $a_{\rm c}/a_{\rm b} \gtrsim 2.12$, whereas actually $a_{\rm c}/a_{\rm b} \approx 2.49$.

Hence, residence in a strong mean-motion resonance is not necessarily required to stabilise the system. Nevertheless, the system's proximity to a strong mean-motion resonance is of interest, particularly in context of the entire exoplanet population. Fig. 5 of \cite{weietal2022} illustrates a statistically significant asymmetry in the population of two-planet pairs which reside just interior versus just exterior to the strongest (first-order) mean-motion resonances, first noted in \citet{Fabrycky2014}. The observed population of $4$:$1$ planetary pairs might not yet be high enough for a $4$:$1$ asymmetry to be detectable. In this respect, the TOI-1052 system might provide a valuable data point, although it cannot be excluded that another planet lies between the two detected planets.

In order to explore the system's proximity to resonance, we employ the semianalytic libration width prescription of \cite{galetal2021}. This prescription effectively computes bounds within which mean-motion resonant behaviour is possible through a numerical procedure mixed in with analytical theory.

\begin{figure}
    \includegraphics[scale=1.0]{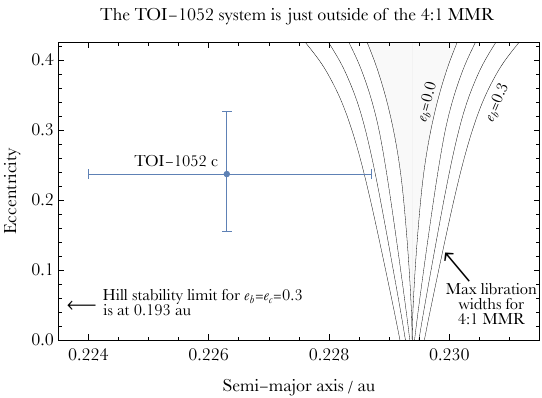}
    \caption{Proximity of the two planets in the TOI-1052 to the $4$:$1$ mean motion resonance. The four pairs of curves are libration widths for this resonance. These curves, moving outwards, correspond to $e_{\rm b} = 0.0, 0.1, 0.2, 0.3$. The planet TOI-1052~c is nearly outside all of these curves, adding to the asymmetry seen around mean-motion commensurabilities in the exoplanet population. This system is also Hill stable, with the critical limit off the scale of the plot. 
    \label{fig:MMR}}
\end{figure}

We plot the libration width curves for four cases ($e_{\rm b} = 0.0, 0.1, 0.2, 0.3$) in Fig. \ref{fig:MMR} by using an eccentricity resolution of 0.025 in the numerical integration. Superimposed are the uncertainties for the current location of TOI-1052~c. Comparing these uncertainties with the libration width locations indicates that the TOI-1052 system is definitely not in resonance (at least a third-, second- or first-order resonance), and resides just interior to the $4$:$1$ resonance. The system's close proximity to resonance is characteristic of many exoplanetary systems, although the proximity to a relatively high-order resonance is noteworthy. Proximity to resonance can be a marker for the differential migration rates of planets in their nascent protoplanetary disc, although to date this has primarily been investigated in depth for first-order resonances \citep{Huang2023}. 

\subsection{Internal structure}
\begin{figure}
  \centering
  \begin{tabular}{@{}cc@{}}
    \includegraphics[scale=0.5]{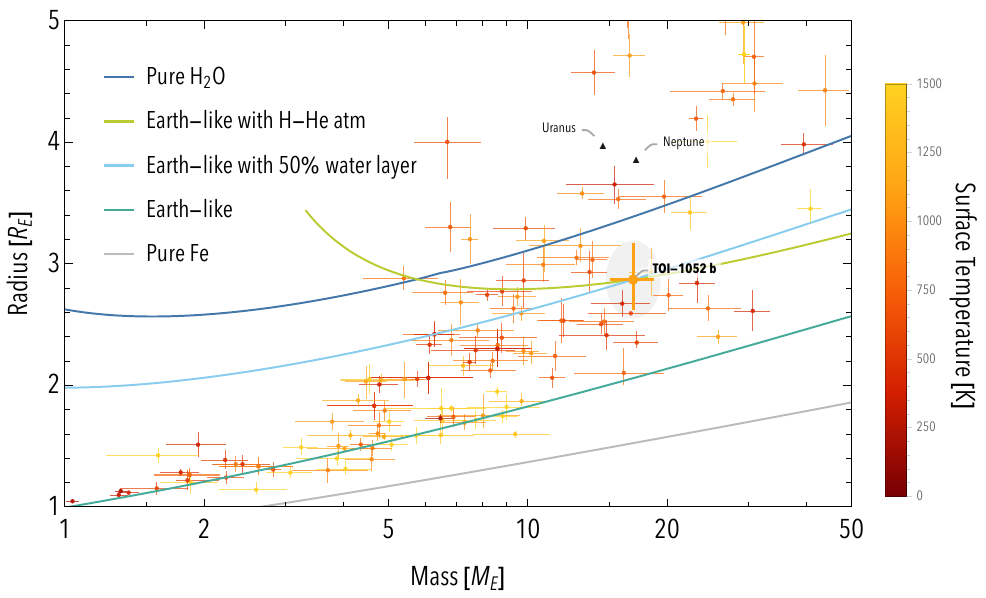}
    \end{tabular}
    \caption{Mass--radius diagram showing various internal structure lines from our model labelled in the legend. TOI-1052b can be explained by an Earth-like composition with either 50\% water or a H-He atmosphere as described in the text.}
    \label{fig:m-r-relation}
\end{figure}

TOI-1052 b is similar to Uranus or Neptune in mass, but has a considerably smaller radius and therefore a denser interior. Fig. \ref{fig:m-r-relation}, which shows the Mass-Radius relation, demonstrates that TOI-1052 b is located between the water line and Earth-like compositional line, suggesting a significant fraction of  refractory materials. In comparison, both Uranus and Neptune are located above the pure-water line. 

For TOI-1052b, two limiting cases come to mind: a planet with a rocky interior and a substantial hydrosphere and a refractory-rich planet with a primordial H-He atmosphere. We investigate these two scenarios with a layered interior model, consisting of up to four layers: a H-He atmosphere, a water layer, a silicate mantle, and an iron core \citep[see][]{Dorn_2017}. Using the inferred age and elemental abundances of TOI-1052, we solve the standard structure equations for two models: 1.) a model where we assume that TOI-1052 b contains no water (\textit{no-water model}) and 2.) a model where we conversely assume that TOI-1052 b contains no H-He atmosphere (\textit{no-atmosphere model}). We put no constraints on  the compositions, i.e., the elemental ratios, of the other layers. As a result, the iron-to-rock ratio can take any value. 
For both models, we apply a nested sampling algorithm \citep{Buchner_2014} to explore the permitted parameter ranges that reproduce the measured masses and radii of TOI-1052 b. 


We find that the no-water model favors a core-to-mantle mass fraction of nearly unity: $0.96 \pm 0.17$ with a H-He envelope of $2^{+1.4}_{-0.8}~\%$. In the case of the no-atmosphere model, while the core-to-mantle mass fraction is poorly constrained ($0.6 \pm 0.5$), this model predicts a water mass fraction of $0.43 \pm 0.12$. The larger uncertainties are caused by the wide range of possible core, mantle, and water layer masses that can reproduce the observed radius and mass compared to the no-water model. Assuming a fixed iron-to-rock ratio, e.g., similar to the host star's elemental ratios, decreases the model's uncertainty significantly.

We note that at high planetary masses, layers might not be as distinct as assumed here \citep[e.g.,][]{Helled2017, Bodenheimer_2018}. Moreover, the atmospheric mass fraction may be underestimated due to pollution of the H-He envelope by heavier elements, leading to further contraction of the atmosphere \citep{Lozovsky_2018}. The interior model also neglects  any water that is dissolved deep in the interior, which could increase the overall water mass fraction \citep{dorn2021hidden}. Nevertheless, while these details could change the exact  values inferred here, it is clear that TOI-1052 b is enriched with refractory materials and any H-He atmosphere is likely to be minimal. 

Additionally, the planet's elemental abundances could differ from its host star, which can change the mantle and the temperature structure. We therefore also considered structure models with varying elemental abundances to investigate this effect. We find that the inferred possible compositions and their error for TOI-1052 b do not change significantly.

\section{Conclusions}

We report the discovery and characterisation of two new planets just outside the 4:1 mean motion resonance in the bright, V=9.5 TOI-1052 system, using \textit{TESS} mission data and HARPS RV measurements. We used high-resolution imaging from the Zorro speckle imaging instrument in order to investigate the presence of any nearby companions and find none within the detector limits. We estimated the projected stellar rotation period to be around 12.8 days from measuring line broadening in the spectra, and derived stellar parameters, chemical abundances and an age estimate to reveal the system in more detail. 

TOI-1052b is a Neptune-mass planet with a sub-Neptune radius, with a potentially eccentric 9.13~d orbit. The planet's density of $3.93^{+1.7}_{-1.3}$~g/cm$^3$ implies a composition denser with more heavy elements than Neptune. Limiting case layered interior models show a degeneracy between a rocky planet with a 2\% H-He atmosphere and a water-rich planet with a water mass fraction of 0.43. 

The companion planet TOI-1052c shows an $M_P\sin i$ of $34.3^{+4.1}_{-3.7}M_{\oplus}$, approx. double the mass of planet b, and orbits on a 35.8d period. Given its presence near the 4:1 mean motion resonance, and the potential eccentricity of both planets, the system provides an interesting case study for dynamical interactions.

\section*{Data Availability}
\textit{TESS} data is accessible via the MAST (Mikulski Archive for
Space Telescopes) portal at https://mast.stsci.edu/portal/
Mashup/Clients/Mast/Portal.html. Imaging
data from Zorro are accessible via the ExoFOP-TESS
archive at https://exofop.ipac.caltech.edu/tess/target.php?id=317060587. The exoplanet modelling code and associated python scripts for parameter analysis and plotting are available
upon reasonable request to the author. Radial velocity data is presented in Table \ref{tab:spec1}.

\section*{Acknowledgements}

Based on observations collected at the La Silla Observatory, ESO(Chile), with the HARPS spectrograph at the 3.6-m telescope for programs $1102.C-0249(F)$, $106.21TJ.001$ and $105.20G9.001$.

DJA is supported by UKRI through the STFC (ST/R00384X/1) and EPSRC (EP/X027562/1). AO and FH are funded by an STFC studentship. Co-funded by the European Union (ERC, FIERCE, 101052347). Views and opinions expressed are however those of the authors only and do not necessarily reflect those of the European Union or the European Research Council. Neither the European Union nor the granting authority can be held responsible for them. This work was supported by FCT - Fundação para a Ciência e a Tecnologia through national funds and by FEDER through COMPETE2020 - Programa Operacional Competitividade e Internacionalização by these grants: UIDB/04434/2020; UIDP/04434/2020. VA acknowledges the support from FCT through the following grant:  2022.06962.PTDC. SGS acknowledges the support from FCT through Investigador FCT contract nr. CEECIND/00826/2018 and POPH/FSE (EC). EDM. acknowledges the support from FCT through Investigador FCT contract nr. 2021.01294.CEECIND. HK and RH carried out this work within the framework of the NCCR PlanetS supported by the Swiss National Science Foundation under grants 51NF40\_182901 and 51NF40\_205606. CD acknowledges support from the Swiss National Science Foundation under grant PZ00P2\_174028. This project has received funding from the European Research Council (ERC) under the European Union’s Horizon 2020 research and innovation programme (grant agreement SCORE No 851555). J.L-B. is partly funded by grants LCF/BQ/PI20/11760023, Ram\'on y Cajal fellowship with code RYC2021-031640-I, and the Spanish MCIN/AEI/10.13039/501100011033 grant PID2019-107061GB-C61. SH acknowledges CNES funding through the grant 837319.

This work made use of \texttt{tpfplotter} by J. Lillo-Box (publicly available in www.github.com/jlillo/tpfplotter), which also made use of the python packages \texttt{astropy}, \texttt{lightkurve}, \texttt{matplotlib} and \texttt{numpy}. 

Funding for the TESS mission is provided by NASA's Science Mission Directorate. We acknowledge the use of public TESS data from pipelines at the TESS Science Office and at the TESS Science Processing Operations Center. Resources supporting this work were provided by the NASA High-End Computing (HEC) Program through the NASA Advanced Supercomputing (NAS) Division at Ames Research Center for the production of the SPOC data products. This research has made use of the Exoplanet Follow-up Observation Program website, which is operated by the California Institute of Technology, under contract with the National Aeronautics and Space Administration under the Exoplanet Exploration Program.

Some of the observations in this paper made use of the High-Resolution Imaging instrument Zorro and were obtained under Gemini LLP Proposal Number: GN/S-2021A-LP-105. Zorro was funded by the NASA Exoplanet Exploration Program and built at the NASA Ames Research Center by Steve B. Howell, Nic Scott, Elliott P. Horch, and Emmett Quigley. Zorro was mounted on the Gemini South telescope of the international Gemini Observatory, a program of NSF’s OIR Lab, which is managed by the Association of Universities for Research in Astronomy (AURA) under a cooperative agreement with the National Science Foundation. on behalf of the Gemini partnership: the National Science Foundation (United States), National Research Council (Canada), Agencia Nacional de Investigación y Desarrollo (Chile), Ministerio de Ciencia, Tecnología e Innovación (Argentina), Ministério da Ciência, Tecnologia, Inovações e Comunicações (Brazil), and Korea Astronomy and Space Science Institute (Republic of Korea).

This work has made use of data from the European Space Agency (ESA) mission
{\it Gaia} (\url{https://www.cosmos.esa.int/gaia}), processed by the {\it Gaia}
Data Processing and Analysis Consortium (DPAC,
\url{https://www.cosmos.esa.int/web/gaia/dpac/consortium}). Funding for the DPAC
has been provided by national institutions, in particular the institutions
participating in the {\it Gaia} Multilateral Agreement.

\bibliographystyle{mnras}
\bibliography{bibliography}

\begin{thebibliography}{}
\makeatletter
\relax
\def\mn@urlcharsother{\let\do\@makeother \do\$\do\&\do\#\do\^\do\_\do\%\do\~}
\def\mn@doi{\begingroup\mn@urlcharsother \@ifnextchar [ {\mn@doi@}
  {\mn@doi@[]}}
\def\mn@doi@[#1]#2{\def\@tempa{#1}\ifx\@tempa\@empty \href
  {http://dx.doi.org/#2} {doi:#2}\else \href {http://dx.doi.org/#2} {#1}\fi
  \endgroup}
\def\mn@eprint#1#2{\mn@eprint@#1:#2::\@nil}
\def\mn@eprint@arXiv#1{\href {http://arxiv.org/abs/#1} {{\tt arXiv:#1}}}
\def\mn@eprint@dblp#1{\href {http://dblp.uni-trier.de/rec/bibtex/#1.xml}
  {dblp:#1}}
\def\mn@eprint@#1:#2:#3:#4\@nil{\def\@tempa {#1}\def\@tempb {#2}\def\@tempc
  {#3}\ifx \@tempc \@empty \let \@tempc \@tempb \let \@tempb \@tempa \fi \ifx
  \@tempb \@empty \def\@tempb {arXiv}\fi \@ifundefined
  {mn@eprint@\@tempb}{\@tempb:\@tempc}{\expandafter \expandafter \csname
  mn@eprint@\@tempb\endcsname \expandafter{\@tempc}}}

\bibitem[\protect\citeauthoryear{{Adibekyan}, {Sousa}, {Santos}, {Delgado
  Mena}, {Gonz{\'a}lez Hern{\'a}ndez}, {Israelian}, {Mayor}  \&
  {Khachatryan}}{{Adibekyan} et~al.}{2012}]{Adibekyan-12}
{Adibekyan} V.~Z.,  {Sousa} S.~G.,  {Santos} N.~C.,  {Delgado Mena} E.,
  {Gonz{\'a}lez Hern{\'a}ndez} J.~I.,  {Israelian} G.,  {Mayor} M.,
  {Khachatryan} G.,  2012, \mn@doi [\aap] {10.1051/0004-6361/201219401}, \href
  {http://adsabs.harvard.edu/abs/2012A%26A...545A..32A} {545, A32}

\bibitem[\protect\citeauthoryear{{Adibekyan} et~al.,}{{Adibekyan}
  et~al.}{2015}]{Adibekyan-15}
{Adibekyan} V.,  et~al., 2015, \mn@doi [\aap] {10.1051/0004-6361/201527120},
  \href {http://adsabs.harvard.edu/abs/2015A%26A...583A..94A} {583, A94}

\bibitem[\protect\citeauthoryear{{Akeson} et~al.,}{{Akeson}
  et~al.}{2013}]{Akeson-13}
{Akeson} R.~L.,  et~al., 2013, \mn@doi [\pasp] {10.1086/672273}, \href
  {https://ui.adsabs.harvard.edu/abs/2013PASP..125..989A} {125, 989}

\bibitem[\protect\citeauthoryear{{Alibert}, {Mordasini}, {Benz}  \&
  {Naef}}{{Alibert} et~al.}{2010}]{Alibert-10}
{Alibert} Y.,  {Mordasini} C.,  {Benz} W.,   {Naef} D.,  2010, in
  {Go{\.z}dziewski} K.,  {Niedzielski} A.,   {Schneider} J.,  eds,  EAS
  Publications Series Vol. 42, EAS Publications Series. pp 209--225,
  \mn@doi{10.1051/eas/1042024}

\bibitem[\protect\citeauthoryear{{Alibert}, {thiabaud}, {marboeuf}, {swoboda},
  {benz}, {mezger}  \& {leya}}{{Alibert} et~al.}{2015}]{Alibert-15}
{Alibert} Y.,  {thiabaud} a.,  {marboeuf} u.,  {swoboda} d.,  {benz} w.,
  {mezger} k.,   {leya} i.,  2015, in AAS/Division for Extreme Solar Systems
  Abstracts. p. 115.02

\bibitem[\protect\citeauthoryear{{Aller}, {Lillo-Box}, {Jones}, {Miranda}  \&
  {Barcel{\'o} Forteza}}{{Aller} et~al.}{2020}]{2020A&A...635A.128A}
{Aller} A.,  {Lillo-Box} J.,  {Jones} D.,  {Miranda} L.~F.,   {Barcel{\'o}
  Forteza} S.,  2020, \mn@doi [\aap] {10.1051/0004-6361/201937118}, \href
  {https://ui.adsabs.harvard.edu/abs/2020A&A...635A.128A} {635, A128}

\bibitem[\protect\citeauthoryear{{Armstrong} et~al.,}{{Armstrong}
  et~al.}{2020}]{Armstrong2020}
{Armstrong} D.~J.,  et~al., 2020, \mn@doi [\nat] {10.1038/s41586-020-2421-7},
  \href {https://ui.adsabs.harvard.edu/abs/2020Natur.583...39A} {583, 39}

\bibitem[\protect\citeauthoryear{{Astropy Collaboration} et~al.,}{{Astropy
  Collaboration} et~al.}{2022}]{Astropy3_2022}
{Astropy Collaboration} et~al., 2022, \mn@doi [\apj]
  {10.3847/1538-4357/ac7c74}, \href
  {https://ui.adsabs.harvard.edu/abs/2022ApJ...935..167A} {935, 167}

\bibitem[\protect\citeauthoryear{{Baranne} et~al.,}{{Baranne}
  et~al.}{1996}]{Baranne-96}
{Baranne} A.,  et~al., 1996, \aaps, \href
  {https://ui.adsabs.harvard.edu/abs/1996A&AS..119..373B} {119, 373}

\bibitem[\protect\citeauthoryear{{Beaug{\'e}} \& {Nesvorn{\'y}}}{{Beaug{\'e}}
  \& {Nesvorn{\'y}}}{2013}]{Beauge-13}
{Beaug{\'e}} C.,  {Nesvorn{\'y}} D.,  2013, \mn@doi [\apj]
  {10.1088/0004-637X/763/1/12}, \href
  {https://ui.adsabs.harvard.edu/abs/2013ApJ...763...12B} {763, 12}

\bibitem[\protect\citeauthoryear{{Bertran de Lis}, {Delgado Mena}, {Adibekyan},
  {Santos}  \& {Sousa}}{{Bertran de Lis} et~al.}{2015}]{Bertrandelis-15}
{Bertran de Lis} S.,  {Delgado Mena} E.,  {Adibekyan} V.~Z.,  {Santos} N.~C.,
  {Sousa} S.~G.,  2015, \mn@doi [\aap] {10.1051/0004-6361/201424633}, \href
  {https://ui.adsabs.harvard.edu/abs/2015A&A...576A..89B} {576, A89}

\bibitem[\protect\citeauthoryear{Bodenheimer, Stevenson, Lissauer  \&
  D’Angelo}{Bodenheimer et~al.}{2018}]{Bodenheimer_2018}
Bodenheimer P.,  Stevenson D.~J.,  Lissauer J.~J.,   D’Angelo G.,  2018,
  \mn@doi [The Astrophysical Journal] {10.3847/1538-4357/aae928}, 868, 138

\bibitem[\protect\citeauthoryear{{Bou{\'e}}, {Figueira}, {Correia}  \&
  {Santos}}{{Bou{\'e}} et~al.}{2012}]{Boue2012}
{Bou{\'e}} G.,  {Figueira} P.,  {Correia} A.~C.~M.,   {Santos} N.~C.,  2012,
  \mn@doi [\aap] {10.1051/0004-6361/201118084}, \href
  {https://ui.adsabs.harvard.edu/abs/2012A&A...537L...3B} {537, L3}

\bibitem[\protect\citeauthoryear{Bourque et~al.,}{Bourque
  et~al.}{2021}]{matthew_bourque_2021}
Bourque M.,  et~al., 2021, The Exoplanet Characterization Toolkit (ExoCTK),
  \mn@doi{10.5281/zenodo.4556063}, \url
  {https://doi.org/10.5281/zenodo.4556063}

\bibitem[\protect\citeauthoryear{Brooks \& Gelman}{Brooks \&
  Gelman}{1998}]{GRstat1998}
Brooks S.~P.,  Gelman A.,  1998, \mn@doi [Journal of Computational and
  Graphical Statistics] {10.1080/10618600.1998.10474787}, 7, 434

\bibitem[\protect\citeauthoryear{{Buchner} et~al.,}{{Buchner}
  et~al.}{2014}]{Buchner_2014}
{Buchner} J.,  et~al., 2014, \mn@doi [\aap] {10.1051/0004-6361/201322971},
  \href {https://ui.adsabs.harvard.edu/abs/2014A&A...564A.125B} {564, A125}

\bibitem[\protect\citeauthoryear{{Costa Silva}, {Delgado Mena}  \&
  {Tsantaki}}{{Costa Silva} et~al.}{2020}]{CostaSilva-20}
{Costa Silva} A.~R.,  {Delgado Mena} E.,   {Tsantaki} M.,  2020, \mn@doi [\aap]
  {10.1051/0004-6361/201936523}, \href
  {https://ui.adsabs.harvard.edu/abs/2020A&A...634A.136C} {634, A136}

\bibitem[\protect\citeauthoryear{{Delgado Mena}, {Tsantaki}, {Adibekyan},
  {Sousa}, {Santos}, {Gonz{\'a}lez Hern{\'a}ndez}  \& {Israelian}}{{Delgado
  Mena} et~al.}{2017}]{DelgadoMena-17}
{Delgado Mena} E.,  {Tsantaki} M.,  {Adibekyan} V.~Z.,  {Sousa} S.~G.,
  {Santos} N.~C.,  {Gonz{\'a}lez Hern{\'a}ndez} J.~I.,   {Israelian} G.,  2017,
  \mn@doi [\aap] {10.1051/0004-6361/201730535}, \href
  {http://adsabs.harvard.edu/abs/2017A%26A...606A..94D} {606, A94}

\bibitem[\protect\citeauthoryear{{Delgado Mena} et~al.,}{{Delgado Mena}
  et~al.}{2019}]{DelgadoMena2019}
{Delgado Mena} E.,  et~al., 2019, \mn@doi [\aap] {10.1051/0004-6361/201834783},
  \href {https://ui.adsabs.harvard.edu/abs/2019A&A...624A..78D} {624, A78}

\bibitem[\protect\citeauthoryear{{Delgado Mena}, {Adibekyan}, {Santos},
  {Tsantaki}, {Gonz{\'a}lez Hern{\'a}ndez}, {Sousa}  \& {Bertr{\'a}n de
  Lis}}{{Delgado Mena} et~al.}{2021}]{DelgadoMena-21}
{Delgado Mena} E.,  {Adibekyan} V.,  {Santos} N.~C.,  {Tsantaki} M.,
  {Gonz{\'a}lez Hern{\'a}ndez} J.~I.,  {Sousa} S.~G.,   {Bertr{\'a}n de Lis}
  S.,  2021, \mn@doi [\aap] {10.1051/0004-6361/202141588}, \href
  {https://ui.adsabs.harvard.edu/abs/2021A&A...655A..99D} {655, A99}

\bibitem[\protect\citeauthoryear{{Donnison}}{{Donnison}}{2006}]{donnison2006}
{Donnison} J.~R.,  2006, \mn@doi [\mnras] {10.1111/j.1365-2966.2006.10372.x},
  \href {https://ui.adsabs.harvard.edu/abs/2006MNRAS.369.1267D} {369, 1267}

\bibitem[\protect\citeauthoryear{{Donnison}}{{Donnison}}{2011}]{donnison2011}
{Donnison} J.~R.,  2011, \mn@doi [\mnras] {10.1111/j.1365-2966.2011.18720.x},
  \href {https://ui.adsabs.harvard.edu/abs/2011MNRAS.415..470D} {415, 470}

\bibitem[\protect\citeauthoryear{Dorn \& Lichtenberg}{Dorn \&
  Lichtenberg}{2021}]{dorn2021hidden}
Dorn C.,  Lichtenberg T.,  2021, The Astrophysical Journal Letters, 922, L4

\bibitem[\protect\citeauthoryear{{Dorn}, {Venturini}, {Khan}, {Heng},
  {Alibert}, {Helled}, {Rivoldini}  \& {Benz}}{{Dorn} et~al.}{2017}]{Dorn_2017}
{Dorn} C.,  {Venturini} J.,  {Khan} A.,  {Heng} K.,  {Alibert} Y.,  {Helled}
  R.,  {Rivoldini} A.,   {Benz} W.,  2017, \mn@doi [\aap]
  {10.1051/0004-6361/201628708}, \href
  {https://ui.adsabs.harvard.edu/abs/2017A&A...597A..37D} {597, A37}

\bibitem[\protect\citeauthoryear{{Doyle}, {Davies}, {Smalley}, {Chaplin}  \&
  {Elsworth}}{{Doyle} et~al.}{2014}]{Doyle2014}
{Doyle} A.~P.,  {Davies} G.~R.,  {Smalley} B.,  {Chaplin} W.~J.,   {Elsworth}
  Y.,  2014, \mn@doi [\mnras] {10.1093/mnras/stu1692}, \href
  {https://ui.adsabs.harvard.edu/abs/2014MNRAS.444.3592D} {444, 3592}

\bibitem[\protect\citeauthoryear{{Fabrycky} et~al.,}{{Fabrycky}
  et~al.}{2014}]{Fabrycky2014}
{Fabrycky} D.~C.,  et~al., 2014, \mn@doi [\apj] {10.1088/0004-637X/790/2/146},
  \href {https://ui.adsabs.harvard.edu/abs/2014ApJ...790..146F} {790, 146}

\bibitem[\protect\citeauthoryear{{Foreman-Mackey} et~al.,}{{Foreman-Mackey}
  et~al.}{2021}]{exoplanet:joss}
{Foreman-Mackey} D.,  et~al., 2021, arXiv e-prints, \href
  {https://ui.adsabs.harvard.edu/abs/2021arXiv210501994F} {p. arXiv:2105.01994}

\bibitem[\protect\citeauthoryear{{Fouesneau} et~al.,}{{Fouesneau}
  et~al.}{2022}]{Gaiastellar2022}
{Fouesneau} M.,  et~al., 2022, \mn@doi [arXiv e-prints]
  {10.48550/arXiv.2206.05992}, \href
  {https://ui.adsabs.harvard.edu/abs/2022arXiv220605992F} {p. arXiv:2206.05992}

\bibitem[\protect\citeauthoryear{{Gaia Collaboration} et~al.,}{{Gaia
  Collaboration} et~al.}{2016}]{GaiaMission2016}
{Gaia Collaboration} et~al., 2016, \mn@doi [\aap]
  {10.1051/0004-6361/201629272}, \href
  {https://ui.adsabs.harvard.edu/abs/2016A&A...595A...1G} {595, A1}

\bibitem[\protect\citeauthoryear{{Gaia Collaboration} et~al.,}{{Gaia
  Collaboration} et~al.}{2022}]{GaiaDR32022}
{Gaia Collaboration} et~al., 2022, \mn@doi [arXiv e-prints]
  {10.48550/arXiv.2208.00211}, \href
  {https://ui.adsabs.harvard.edu/abs/2022arXiv220800211G} {p. arXiv:2208.00211}

\bibitem[\protect\citeauthoryear{{Gallardo}, {Beaug{\'e}}  \&
  {Giuppone}}{{Gallardo} et~al.}{2021}]{galetal2021}
{Gallardo} T.,  {Beaug{\'e}} C.,   {Giuppone} C.~A.,  2021, \mn@doi [\aap]
  {10.1051/0004-6361/202039764}, \href
  {https://ui.adsabs.harvard.edu/abs/2021A&A...646A.148G} {646, A148}

\bibitem[\protect\citeauthoryear{{Guerrero} et~al.,}{{Guerrero}
  et~al.}{2021}]{Guerrero2021}
{Guerrero} N.~M.,  et~al., 2021, \mn@doi [\apjs] {10.3847/1538-4365/abefe1},
  \href {https://ui.adsabs.harvard.edu/abs/2021ApJS..254...39G} {254, 39}

\bibitem[\protect\citeauthoryear{{Hara}, {Bou{\'e}}, {Laskar}  \&
  {Correia}}{{Hara} et~al.}{2017}]{Hara-17}
{Hara} N.~C.,  {Bou{\'e}} G.,  {Laskar} J.,   {Correia} A.~C.~M.,  2017,
  \mn@doi [\mnras] {10.1093/mnras/stw2261}, \href
  {https://ui.adsabs.harvard.edu/abs/2017MNRAS.464.1220H} {464, 1220}

\bibitem[\protect\citeauthoryear{{Hara} et~al.,}{{Hara} et~al.}{2020}]{Hara-20}
{Hara} N.~C.,  et~al., 2020, \mn@doi [\aap] {10.1051/0004-6361/201937254},
  \href {https://ui.adsabs.harvard.edu/abs/2020A&A...636L...6H} {636, L6}

\bibitem[\protect\citeauthoryear{{Helled} \& {Stevenson}}{{Helled} \&
  {Stevenson}}{2017}]{Helled2017}
{Helled} R.,  {Stevenson} D.,  2017, \mn@doi [\apjl]
  {10.3847/2041-8213/aa6d08}, \href
  {https://ui.adsabs.harvard.edu/abs/2017ApJ...840L...4H} {840, L4}

\bibitem[\protect\citeauthoryear{{Hippke} \& {Heller}}{{Hippke} \&
  {Heller}}{2019}]{2019A&A...623A..39H}
{Hippke} M.,  {Heller} R.,  2019, \mn@doi [\aap] {10.1051/0004-6361/201834672},
  \href {https://ui.adsabs.harvard.edu/\#abs/2019A&A...623A..39H} {623, A39}

\bibitem[\protect\citeauthoryear{{H{\o}g} et~al.,}{{H{\o}g}
  et~al.}{2000}]{Tycho}
{H{\o}g} E.,  et~al., 2000, \aap, \href
  {https://ui.adsabs.harvard.edu/abs/2000A&A...355L..27H} {355, L27}

\bibitem[\protect\citeauthoryear{{Hojjatpanah} et~al.,}{{Hojjatpanah}
  et~al.}{2019}]{2019A&A...629A..80H}
{Hojjatpanah} S.,  et~al., 2019, \mn@doi [\aap] {10.1051/0004-6361/201834729},
  \href {https://ui.adsabs.harvard.edu/abs/2019A&A...629A..80H} {629, A80}

\bibitem[\protect\citeauthoryear{{Houk} \& {Cowley}}{{Houk} \&
  {Cowley}}{1975}]{1975mcts.book.....H}
{Houk} N.,  {Cowley} A.~P.,  1975, {University of Michigan Catalogue of
  two-dimensional spectral types for the HD stars. Volume I. Declinations -90\_
  to -53\_{\textflorin}0.}

\bibitem[\protect\citeauthoryear{{Howell} \& {Furlan}}{{Howell} \&
  {Furlan}}{2022}]{Howell2022}
{Howell} S.~B.,  {Furlan} E.,  2022, \mn@doi [Frontiers in Astronomy and Space
  Sciences] {10.3389/fspas.2022.871163}, \href
  {https://ui.adsabs.harvard.edu/abs/2022FrASS...9.1163H} {9, 871163}

\bibitem[\protect\citeauthoryear{{Howell}, {Everett}, {Sherry}, {Horch}  \&
  {Ciardi}}{{Howell} et~al.}{2011}]{Howell2011}
{Howell} S.~B.,  {Everett} M.~E.,  {Sherry} W.,  {Horch} E.,   {Ciardi} D.~R.,
  2011, \mn@doi [\aj] {10.1088/0004-6256/142/1/19}, \href
  {https://ui.adsabs.harvard.edu/abs/2011AJ....142...19H} {142, 19}

\bibitem[\protect\citeauthoryear{Hoyer et~al.,}{Hoyer
  et~al.}{2021}]{SHoyer2021}
Hoyer S.,  et~al., 2021, \mn@doi [Monthly Notices of the Royal Astronomical
  Society] {10.1093/mnras/stab1427}, 505, 3361

\bibitem[\protect\citeauthoryear{{Huang} \& {Ormel}}{{Huang} \&
  {Ormel}}{2023}]{Huang2023}
{Huang} S.,  {Ormel} C.~W.,  2023, \mn@doi [\mnras] {10.1093/mnras/stad1032},
  \href {https://ui.adsabs.harvard.edu/abs/2023MNRAS.522..828H} {522, 828}

\bibitem[\protect\citeauthoryear{{Jenkins}}{{Jenkins}}{2002}]{Jenkins2002}
{Jenkins} J.~M.,  2002, \mn@doi [\apj] {10.1086/341136}, \href
  {https://ui.adsabs.harvard.edu/abs/2002ApJ...575..493J} {575, 493}

\bibitem[\protect\citeauthoryear{{Jenkins} et~al.,}{{Jenkins}
  et~al.}{2010a}]{Jenkins2010a}
{Jenkins} J.~M.,  et~al., 2010a, \mn@doi [\apjl] {10.1088/2041-8205/713/2/L87},
  \href {https://ui.adsabs.harvard.edu/abs/2010ApJ...713L..87J} {713, L87}

\bibitem[\protect\citeauthoryear{{Jenkins} et~al.,}{{Jenkins}
  et~al.}{2010b}]{Jenkins2010b}
{Jenkins} J.~M.,  et~al., 2010b, in {Radziwill} N.~M.,  {Bridger} A.,  eds,
  Society of Photo-Optical Instrumentation Engineers (SPIE) Conference Series
  Vol. 7740, Software and Cyberinfrastructure for Astronomy. p. 77400D,
  \mn@doi{10.1117/12.856764}

\bibitem[\protect\citeauthoryear{{Jenkins} et~al.,}{{Jenkins}
  et~al.}{2016}]{Jenkins-16}
{Jenkins} J.~M.,  et~al., 2016, in Software and Cyberinfrastructure for
  Astronomy IV. p. 99133E, \mn@doi{10.1117/12.2233418}

\bibitem[\protect\citeauthoryear{{Jenkins}, {Tenenbaum}, {Seader}, {Burke},
  {McCauliff}, {Smith}, {Twicken}  \& {Chandrasekaran}}{{Jenkins}
  et~al.}{2020}]{Jenkins2020}
{Jenkins} J.~M.,  {Tenenbaum} P.,  {Seader} S.,  {Burke} C.~J.,  {McCauliff}
  S.~D.,  {Smith} J.~C.,  {Twicken} J.~D.,   {Chandrasekaran} H.,  2020,
  {Kepler Data Processing Handbook: Transiting Planet Search}, Kepler Science
  Document KSCI-19081-003, id. 9. Edited by Jon M. Jenkins.

\bibitem[\protect\citeauthoryear{{Kipping}}{{Kipping}}{2013}]{kipping2013}
{Kipping} D.~M.,  2013, \mn@doi [\mnras] {10.1093/mnras/stt1435}, \href
  {https://ui.adsabs.harvard.edu/abs/2013MNRAS.435.2152K} {435, 2152}

\bibitem[\protect\citeauthoryear{{Kurucz}}{{Kurucz}}{1993}]{Kurucz-93}
{Kurucz} R.~L.,  1993, {SYNTHE spectrum synthesis programs and line data}

\bibitem[\protect\citeauthoryear{{Li}, {Tenenbaum}, {Twicken}, {Burke},
  {Jenkins}, {Quintana}, {Rowe}  \& {Seader}}{{Li} et~al.}{2019}]{Li-19}
{Li} J.,  {Tenenbaum} P.,  {Twicken} J.~D.,  {Burke} C.~J.,  {Jenkins} J.~M.,
  {Quintana} E.~V.,  {Rowe} J.~F.,   {Seader} S.~E.,  2019, \mn@doi [\pasp]
  {10.1088/1538-3873/aaf44d}, \href
  {https://ui.adsabs.harvard.edu/abs/2019PASP..131b4506L} {131, 024506}

\bibitem[\protect\citeauthoryear{{Lomb}}{{Lomb}}{1976}]{Lomb1976}
{Lomb} N.~R.,  1976, \mn@doi [\apss] {10.1007/BF00648343}, \href
  {https://ui.adsabs.harvard.edu/abs/1976Ap&SS..39..447L} {39, 447}

\bibitem[\protect\citeauthoryear{{Lozovsky}, {Helled}, {Dorn}  \&
  {Venturini}}{{Lozovsky} et~al.}{2018}]{Lozovsky_2018}
{Lozovsky} M.,  {Helled} R.,  {Dorn} C.,   {Venturini} J.,  2018, \mn@doi
  [\apj] {10.3847/1538-4357/aadd09}, \href
  {https://ui.adsabs.harvard.edu/abs/2018ApJ...866...49L} {866, 49}

\bibitem[\protect\citeauthoryear{{Luger}, {Agol}, {Foreman-Mackey}, {Fleming},
  {Lustig-Yaeger}  \& {Deitrick}}{{Luger} et~al.}{2019}]{starryLuger2019}
{Luger} R.,  {Agol} E.,  {Foreman-Mackey} D.,  {Fleming} D.~P.,
  {Lustig-Yaeger} J.,   {Deitrick} R.,  2019, \mn@doi [\aj]
  {10.3847/1538-3881/aae8e5}, \href
  {https://ui.adsabs.harvard.edu/abs/2019AJ....157...64L} {157, 64}

\bibitem[\protect\citeauthoryear{Maldonado et~al.,}{Maldonado
  et~al.}{2017}]{maldonado2017hades}
Maldonado J.,  et~al., 2017, Astronomy \& Astrophysics, 598, A27

\bibitem[\protect\citeauthoryear{{Mayor} et~al.,}{{Mayor}
  et~al.}{2003}]{Mayor-13}
{Mayor} M.,  et~al., 2003, The Messenger, \href
  {http://adsabs.harvard.edu/abs/2003Msngr.114...20M} {114, 20}

\bibitem[\protect\citeauthoryear{{Mazeh}, {Holczer}  \& {Faigler}}{{Mazeh}
  et~al.}{2016}]{Mazeh2016}
{Mazeh} T.,  {Holczer} T.,   {Faigler} S.,  2016, \mn@doi [\aap]
  {10.1051/0004-6361/201528065}, \href
  {https://ui.adsabs.harvard.edu/abs/2016A&A...589A..75M} {589, A75}

\bibitem[\protect\citeauthoryear{{Nielsen} et~al.,}{{Nielsen}
  et~al.}{2020}]{Nielsen2020}
{Nielsen} L.~D.,  et~al., 2020, \mn@doi [\mnras] {10.1093/mnras/staa197}, \href
  {https://ui.adsabs.harvard.edu/abs/2020MNRAS.492.5399N} {492, 5399}

\bibitem[\protect\citeauthoryear{{Otegi} et~al.,}{{Otegi}
  et~al.}{2021}]{Otegi2021}
{Otegi} J.~F.,  et~al., 2021, \mn@doi [\aap] {10.1051/0004-6361/202040247},
  \href {https://ui.adsabs.harvard.edu/abs/2021A&A...653A.105O} {653, A105}

\bibitem[\protect\citeauthoryear{{Pepe} et~al.,}{{Pepe} et~al.}{2002}]{Pepe-02}
{Pepe} F.,  et~al., 2002, The Messenger, \href
  {http://adsabs.harvard.edu/abs/2002Msngr.110....9P} {110, 9}

\bibitem[\protect\citeauthoryear{{Ricker} et~al.,}{{Ricker}
  et~al.}{2015}]{Ricker-15}
{Ricker} G.~R.,  et~al., 2015, \mn@doi [Journal of Astronomical Telescopes,
  Instruments, and Systems] {10.1117/1.JATIS.1.1.014003}, \href
  {https://ui.adsabs.harvard.edu/abs/2015JATIS...1a4003R} {1, 014003}

\bibitem[\protect\citeauthoryear{{Salvatier}, {Wiecki{\^a}}  \&
  {Fonnesbeck}}{{Salvatier} et~al.}{2016}]{SalvatierPymc3}
{Salvatier} J.,  {Wiecki{\^a}} T.~V.,   {Fonnesbeck} C.,  2016, {PyMC3: Python
  probabilistic programming framework}, Astrophysics Source Code Library,
  record ascl:1610.016 (\mn@eprint {ascl} {1610.016})

\bibitem[\protect\citeauthoryear{{Santos} et~al.,}{{Santos}
  et~al.}{2002}]{2002A&A...392..215S}
{Santos} N.~C.,  et~al., 2002, \mn@doi [\aap] {10.1051/0004-6361:20020876},
  \href {https://ui.adsabs.harvard.edu/abs/2002A&A...392..215S} {392, 215}

\bibitem[\protect\citeauthoryear{{Santos} et~al.,}{{Santos}
  et~al.}{2013}]{Santos-13}
{Santos} N.~C.,  et~al., 2013, \mn@doi [\aap] {10.1051/0004-6361/201321286},
  \href {http://adsabs.harvard.edu/abs/2013A%26A...556A.150S} {556, A150}

\bibitem[\protect\citeauthoryear{{Scargle}}{{Scargle}}{1982}]{Scargle1982}
{Scargle} J.~D.,  1982, \mn@doi [\apj] {10.1086/160554}, \href
  {https://ui.adsabs.harvard.edu/abs/1982ApJ...263..835S} {263, 835}

\bibitem[\protect\citeauthoryear{{Schlegel}, {Finkbeiner}  \&
  {Davis}}{{Schlegel} et~al.}{1998}]{Schlegel:1998}
{Schlegel} D.~J.,  {Finkbeiner} D.~P.,   {Davis} M.,  1998, \mn@doi [\apj]
  {10.1086/305772}, \href
  {https://ui.adsabs.harvard.edu/abs/1998ApJ...500..525S} {500, 525}

\bibitem[\protect\citeauthoryear{{Scott} et~al.,}{{Scott}
  et~al.}{2021}]{Scott2021}
{Scott} N.~J.,  et~al., 2021, \mn@doi [Frontiers in Astronomy and Space
  Sciences] {10.3389/fspas.2021.716560}, \href
  {https://ui.adsabs.harvard.edu/abs/2021FrASS...8..138S} {8, 138}

\bibitem[\protect\citeauthoryear{{Skrutskie} et~al.}{{Skrutskie}
  et~al.}{2006}]{2MASS}
{Skrutskie} M.~F.,  et~al., 2006, \mn@doi [\aj] {10.1086/498708}, \href
  {http://adsabs.harvard.edu/abs/2006AJ....131.1163S} {131, 1163}

\bibitem[\protect\citeauthoryear{{Smith} et~al.,}{{Smith}
  et~al.}{2012}]{Smith-12}
{Smith} J.~C.,  et~al., 2012, \mn@doi [\pasp] {10.1086/667697}, \href
  {http://adsabs.harvard.edu/abs/2012PASP..124.1000S} {124, 1000}

\bibitem[\protect\citeauthoryear{{Sneden}}{{Sneden}}{1973}]{Sneden-73}
{Sneden} C.~A.,  1973, PhD thesis, THE UNIVERSITY OF TEXAS AT AUSTIN.

\bibitem[\protect\citeauthoryear{{Sousa}}{{Sousa}}{2014}]{Sousa-14}
{Sousa} S.~G.,  2014, [arXiv:1407.5817], \href
  {http://adsabs.harvard.edu/abs/2014arXiv1407.5817S} {}

\bibitem[\protect\citeauthoryear{{Sousa}, {Santos}, {Israelian}, {Mayor}  \&
  {Monteiro}}{{Sousa} et~al.}{2007}]{Sousa-07}
{Sousa} S.~G.,  {Santos} N.~C.,  {Israelian} G.,  {Mayor} M.,   {Monteiro}
  M.~J.~P.~F.~G.,  2007, \mn@doi [\aap] {10.1051/0004-6361:20077288}, \href
  {https://ui.adsabs.harvard.edu/abs/2007A&A...469..783S} {469, 783}

\bibitem[\protect\citeauthoryear{{Sousa} et~al.,}{{Sousa}
  et~al.}{2008}]{Sousa-08}
{Sousa} S.~G.,  et~al., 2008, \mn@doi [\aap] {10.1051/0004-6361:200809698},
  \href {https://ui.adsabs.harvard.edu/abs/2008A&A...487..373S} {487, 373}

\bibitem[\protect\citeauthoryear{{Sousa}, {Santos}, {Adibekyan}, {Delgado-Mena}
   \& {Israelian}}{{Sousa} et~al.}{2015}]{Sousa-15}
{Sousa} S.~G.,  {Santos} N.~C.,  {Adibekyan} V.,  {Delgado-Mena} E.,
  {Israelian} G.,  2015, \mn@doi [\aap] {10.1051/0004-6361/201425463}, \href
  {http://adsabs.harvard.edu/abs/2015A%26A...577A..67S} {577, A67}

\bibitem[\protect\citeauthoryear{{Sousa} et~al.,}{{Sousa}
  et~al.}{2021}]{Sousa-21}
{Sousa} S.~G.,  et~al., 2021, \mn@doi [\aap] {10.1051/0004-6361/202141584},
  \href {https://ui.adsabs.harvard.edu/abs/2021A&A...656A..53S} {656, A53}

\bibitem[\protect\citeauthoryear{{Stassun} \& {Torres}}{{Stassun} \&
  {Torres}}{2016}]{Stassun:2016}
{Stassun} K.~G.,  {Torres} G.,  2016, \mn@doi [\aj]
  {10.3847/0004-6256/152/6/180}, \href
  {https://ui.adsabs.harvard.edu/abs/2016AJ....152..180S} {152, 180}

\bibitem[\protect\citeauthoryear{{Stassun} \& {Torres}}{{Stassun} \&
  {Torres}}{2021}]{StassunTorres2021}
{Stassun} K.~G.,  {Torres} G.,  2021, \mn@doi [\apjl]
  {10.3847/2041-8213/abdaad}, \href
  {https://ui.adsabs.harvard.edu/abs/2021ApJ...907L..33S} {907, L33}

\bibitem[\protect\citeauthoryear{{Stassun}, {Collins}  \& {Gaudi}}{{Stassun}
  et~al.}{2017}]{Stassun:2017}
{Stassun} K.~G.,  {Collins} K.~A.,   {Gaudi} B.~S.,  2017, \mn@doi [\aj]
  {10.3847/1538-3881/aa5df3}, \href
  {https://ui.adsabs.harvard.edu/abs/2017AJ....153..136S} {153, 136}

\bibitem[\protect\citeauthoryear{{Stassun}, {Corsaro}, {Pepper}  \&
  {Gaudi}}{{Stassun} et~al.}{2018}]{Stassun:2018}
{Stassun} K.~G.,  {Corsaro} E.,  {Pepper} J.~A.,   {Gaudi} B.~S.,  2018,
  \mn@doi [\aj] {10.3847/1538-3881/aa998a}, \href
  {https://ui.adsabs.harvard.edu/abs/2018AJ....155...22S} {155, 22}

\bibitem[\protect\citeauthoryear{{Stassun} et~al.,}{{Stassun}
  et~al.}{2019}]{Stassun2019TIC}
{Stassun} K.~G.,  et~al., 2019, \mn@doi [\aj] {10.3847/1538-3881/ab3467}, \href
  {https://ui.adsabs.harvard.edu/abs/2019AJ....158..138S} {158, 138}

\bibitem[\protect\citeauthoryear{{Stumpe} et~al.,}{{Stumpe}
  et~al.}{2012}]{Stumpe2012}
{Stumpe} M.~C.,  et~al., 2012, \mn@doi [\pasp] {10.1086/667698}, \href
  {https://ui.adsabs.harvard.edu/abs/2012PASP..124..985S} {124, 985}

\bibitem[\protect\citeauthoryear{{Stumpe}, {Smith}, {Catanzarite}, {Van Cleve},
  {Jenkins}, {Twicken}  \& {Girouard}}{{Stumpe} et~al.}{2014}]{Stumpe2014}
{Stumpe} M.~C.,  {Smith} J.~C.,  {Catanzarite} J.~H.,  {Van Cleve} J.~E.,
  {Jenkins} J.~M.,  {Twicken} J.~D.,   {Girouard} F.~R.,  2014, \mn@doi [\pasp]
  {10.1086/674989}, \href {http://adsabs.harvard.edu/abs/2014PASP..126..100S}
  {126, 100}

\bibitem[\protect\citeauthoryear{{Szab{\'o}} \& {Kiss}}{{Szab{\'o}} \&
  {Kiss}}{2011}]{Szabo-11}
{Szab{\'o}} G.~M.,  {Kiss} L.~L.,  2011, \mn@doi [\apjl]
  {10.1088/2041-8205/727/2/L44}, \href
  {https://ui.adsabs.harvard.edu/abs/2011ApJ...727L..44S} {727, L44}

\bibitem[\protect\citeauthoryear{{Torres}, {Andersen}  \&
  {Gim{\'e}nez}}{{Torres} et~al.}{2010a}]{Torres-2010}
{Torres} G.,  {Andersen} J.,   {Gim{\'e}nez} A.,  2010a, \mn@doi [\aapr]
  {10.1007/s00159-009-0025-1}, \href
  {https://ui.adsabs.harvard.edu/abs/2010A&ARv..18...67T} {18, 67}

\bibitem[\protect\citeauthoryear{{Torres}, {Andersen}  \&
  {Gim{\'e}nez}}{{Torres} et~al.}{2010b}]{Torres:2010}
{Torres} G.,  {Andersen} J.,   {Gim{\'e}nez} A.,  2010b, \mn@doi [\aapr]
  {10.1007/s00159-009-0025-1}, \href
  {https://ui.adsabs.harvard.edu/abs/2010A&ARv..18...67T} {18, 67}

\bibitem[\protect\citeauthoryear{{Twicken}, {Chandrasekaran}, {Jenkins},
  {Gunter}, {Girouard}  \& {Klaus}}{{Twicken} et~al.}{2010}]{Twicken-10}
{Twicken} J.~D.,  {Chandrasekaran} H.,  {Jenkins} J.~M.,  {Gunter} J.~P.,
  {Girouard} F.,   {Klaus} T.~C.,  2010, in Software and Cyberinfrastructure
  for Astronomy. p. 77401U, \mn@doi{10.1117/12.856798}

\bibitem[\protect\citeauthoryear{{Twicken} et~al.,}{{Twicken}
  et~al.}{2018}]{Twicken-18}
{Twicken} J.~D.,  et~al., 2018, \mn@doi [\pasp] {10.1088/1538-3873/aab694},
  \href {http://adsabs.harvard.edu/abs/2018PASP..130f4502T} {130, 064502}

\bibitem[\protect\citeauthoryear{{VanderPlas}}{{VanderPlas}}{2018}]{Vanderplas2018}
{VanderPlas} J.~T.,  2018, \mn@doi [\apjs] {10.3847/1538-4365/aab766}, \href
  {https://ui.adsabs.harvard.edu/abs/2018ApJS..236...16V} {236, 16}

\bibitem[\protect\citeauthoryear{Vehtari, Gelman  \& Gabry}{Vehtari
  et~al.}{2017}]{Vehtari2017}
Vehtari A.,  Gelman A.,   Gabry J.,  2017, \mn@doi [Statistics and Computing]
  {10.1007/s11222-016-9696-4}, 27, 1413

\bibitem[\protect\citeauthoryear{{Veras}, {Mustill}, {Bonsor}  \&
  {Wyatt}}{{Veras} et~al.}{2013}]{veretal2013}
{Veras} D.,  {Mustill} A.~J.,  {Bonsor} A.,   {Wyatt} M.~C.,  2013, \mn@doi
  [\mnras] {10.1093/mnras/stt289}, \href
  {https://ui.adsabs.harvard.edu/abs/2013MNRAS.431.1686V} {431, 1686}

\bibitem[\protect\citeauthoryear{Watanabe}{Watanabe}{2010}]{watanabe10a}
Watanabe S.,  2010, Journal of Machine Learning Research, 11, 3571

\bibitem[\protect\citeauthoryear{{Weiss}, {Millholland}, {Petigura}, {Adams},
  {Batygin}, {Bloch}  \& {Mordasini}}{{Weiss} et~al.}{2022}]{weietal2022}
{Weiss} L.~M.,  {Millholland} S.~C.,  {Petigura} E.~A.,  {Adams} F.~C.,
  {Batygin} K.,  {Bloch} A.~M.,   {Mordasini} C.,  2022, \mn@doi [arXiv
  e-prints] {10.48550/arXiv.2203.10076}, \href
  {https://ui.adsabs.harvard.edu/abs/2022arXiv220310076W} {p. arXiv:2203.10076}

\bibitem[\protect\citeauthoryear{{Wittenmyer}, {Bergmann}, {Horner}, {Clark}
  \& {Kane}}{{Wittenmyer} et~al.}{2019}]{2019MNRAS.484.4230W}
{Wittenmyer} R.~A.,  {Bergmann} C.,  {Horner} J.,  {Clark} J.,   {Kane} S.~R.,
  2019, \mn@doi [\mnras] {10.1093/mnras/stz236}, \href
  {https://ui.adsabs.harvard.edu/abs/2019MNRAS.484.4230W} {484, 4230}

\bibitem[\protect\citeauthoryear{{Wright} et~al.}{{Wright} et~al.}{2010}]{WISE}
{Wright} E.~L.,  et~al., 2010, \mn@doi [\aj] {10.1088/0004-6256/140/6/1868},
  \href {http://adsabs.harvard.edu/abs/2010AJ....140.1868W} {140, 1868}

\bibitem[\protect\citeauthoryear{{da Silva} et~al.,}{{da Silva}
  et~al.}{2006}]{daSilva-06}
{da Silva} L.,  et~al., 2006, \mn@doi [\aap] {10.1051/0004-6361:20065105},
  \href {https://ui.adsabs.harvard.edu/abs/2006A&A...458..609D} {458, 609}

\makeatother
\end{thebibliography}

\appendix
\section{HARPS spectroscopy}

\begin{table*}[ht]
\caption{HARPS spectroscopy.}
\label{tab:spec1}
\scriptsize
\begin{tabular}{lcccccccc}
\hline
\hline
Time & RV & $\sigma_{\rm RV}$  & FWHM & $\sigma_{\rm FWHM}$ & contrast & bis span & $S_{\rm MW}$ & $Ha$ \\
$[\rm{BJD}-2457000]$ & \multicolumn{4}{c}{[m/s]} & \multicolumn{4}{c}{} \\
\hline
\hline

$2358.879$ & $54939.1291$ & $1.7636$ & $9116.2436$ & $1.6357$ & $33.8180$ & $28.8576$ & $0.0939$ & $0.1816$ \\
$2359.889$ & $54942.2963$ & $1.8403$ & $9120.0475$ & $1.6576$ & $33.8104$ & $39.9679$ & $0.0952$ & $0.1776$ \\
$2360.895$ & $54941.4163$ & $1.9099$ & $9116.7254$ & $2.3186$ & $33.7439$ & $28.4139$ & $0.0929$ & $0.1771$ \\
$2361.823$ & $54947.1419$ & $1.9352$ & $9119.1708$ & $1.9099$ & $33.7593$ & $39.5333$ & $0.0945$ & $0.1767$ \\
$2368.779$ & $54936.1522$ & $1.4481$ & $9116.7558$ & $1.7636$ & $33.7817$ & $35.7092$ & $0.0961$ & $0.1808$ \\
$2368.892$ & $54938.6397$ & $1.4729$ & $9112.4784$ & $1.8159$ & $33.7737$ & $35.3361$ & $0.0976$ & $0.1771$ \\
$2369.822$ & $54941.4063$ & $1.7946$ & $9118.8143$ & $1.5630$ & $33.7737$ & $31.2242$ & $0.0950$ & $0.1767$ \\
$2369.901$ & $54943.1930$ & $2.7717$ & $9111.5158$ & $1.5019$ & $33.7755$ & $32.1908$ & $0.0965$ & $0.1773$ \\
$2370.791$ & $54946.6359$ & $2.1584$ & $9119.6451$ & $1.7115$ & $33.8064$ & $30.9872$ & $0.0948$ & $0.1759$ \\
$2370.916$ & $54948.0868$ & $1.7343$ & $9112.1726$ & $2.0355$ & $33.8194$ & $38.4706$ & $0.0926$ & $0.1768$ \\
$2371.810$ & $54947.3120$ & $2.1132$ & $9119.1244$ & $1.5953$ & $33.7795$ & $33.1299$ & $0.0950$ & $0.1812$ \\
$2371.890$ & $54945.3488$ & $2.4833$ & $9123.6366$ & $1.5901$ & $33.8088$ & $34.4833$ & $0.0929$ & $0.1815$ \\
$2372.771$ & $54948.3852$ & $2.9147$ & $9118.4646$ & $1.9352$ & $33.8020$ & $40.1501$ & $0.0922$ & $0.1812$ \\
$2372.896$ & $54946.2718$ & $2.7934$ & $9117.4986$ & $1.6977$ & $33.8151$ & $33.1816$ & $0.0969$ & $0.1818$ \\
$2376.863$ & $54947.9693$ & $1.5630$ & $9119.9399$ & $1.7149$ & $33.7617$ & $41.4132$ & $0.0975$ & $0.1820$ \\
$2378.821$ & $54951.3024$ & $2.0355$ & $9107.8579$ & $1.6013$ & $33.7624$ & $37.8135$ & $0.0971$ & $0.1769$ \\
$2405.829$ & $54938.4662$ & $2.2880$ & $9124.1644$ & $1.8716$ & $33.7738$ & $39.6036$ & $0.0950$ & $0.1820$ \\
$2409.794$ & $54946.2449$ & $1.8159$ & $9102.3543$ & $2.4664$ & $33.8562$ & $25.0392$ & $0.1001$ & $0.1834$ \\
$2412.796$ & $54945.5650$ & $1.7984$ & $9126.1584$ & $1.6767$ & $33.7726$ & $28.3786$ & $0.0964$ & $0.1818$ \\
$2419.683$ & $54954.3765$ & $1.8323$ & $9122.0819$ & $2.1584$ & $33.6817$ & $37.0738$ & $0.0968$ & $0.1827$ \\
$2419.789$ & $54956.1376$ & $1.6042$ & $9113.9300$ & $1.7343$ & $33.7186$ & $37.8870$ & $0.0959$ & $0.1820$ \\
$2421.707$ & $54943.2344$ & $1.4407$ & $9126.5836$ & $2.1132$ & $33.7816$ & $34.4838$ & $0.0946$ & $0.1771$ \\
$2421.806$ & $54947.2486$ & $1.7179$ & $9131.7791$ & $2.4833$ & $33.7808$ & $36.0406$ & $0.0952$ & $0.1770$ \\
$2422.712$ & $54945.2644$ & $2.5884$ & $9132.1542$ & $2.9147$ & $33.7533$ & $35.6213$ & $0.0899$ & $0.1826$ \\
$2422.800$ & $54945.6395$ & $1.9274$ & $9134.2216$ & $2.7934$ & $33.7735$ & $43.7364$ & $0.0956$ & $0.1815$ \\
$2422.935$ & $54942.7393$ & $2.7407$ & $9114.8738$ & $2.5884$ & $33.7706$ & $35.6914$ & $0.0915$ & $0.1817$ \\
$2423.842$ & $54941.5699$ & $1.5777$ & $9116.5795$ & $1.9274$ & $33.7454$ & $37.4534$ & $0.0947$ & $0.1819$ \\
$2423.902$ & $54942.5696$ & $2.0993$ & $9133.7634$ & $2.7407$ & $33.7231$ & $33.5343$ & $0.0921$ & $0.1775$ \\
$2424.785$ & $54948.8373$ & $1.6977$ & $9117.0484$ & $2.0096$ & $33.7747$ & $26.9491$ & $0.0998$ & $0.1783$ \\
$2425.677$ & $54953.2174$ & $2.0096$ & $9119.6364$ & $2.1990$ & $33.7880$ & $32.6183$ & $0.0971$ & $0.1787$ \\
$2425.794$ & $54948.1586$ & $2.1990$ & $9109.2590$ & $1.8323$ & $33.7758$ & $35.9231$ & $0.0987$ & $0.1820$ \\
$2429.803$ & $54944.2671$ & $2.4664$ & $9119.3548$ & $1.6042$ & $33.7349$ & $30.1954$ & $0.0972$ & $0.1804$ \\
$2443.632$ & $54948.4897$ & $1.5019$ & $9133.8345$ & $2.0077$ & $33.7390$ & $38.6013$ & $0.0981$ & $0.1761$ \\
$2443.730$ & $54946.2780$ & $1.7115$ & $9120.9396$ & $1.5777$ & $33.7542$ & $28.5068$ & $0.0941$ & $0.1802$ \\
$2444.647$ & $54947.6991$ & $2.8436$ & $9122.2819$ & $2.0993$ & $33.7997$ & $29.5741$ & $0.0958$ & $0.1811$ \\
$2444.727$ & $54952.9344$ & $3.0458$ & $9109.2394$ & $1.7946$ & $33.7825$ & $27.8161$ & $0.0935$ & $0.1818$ \\
$2449.604$ & $54943.1374$ & $2.0077$ & $9104.2840$ & $2.7717$ & $33.7045$ & $36.4253$ & $0.0930$ & $0.1806$ \\
$2460.746$ & $54941.9471$ & $1.6576$ & $9118.0781$ & $2.2880$ & $33.7604$ & $32.3109$ & $0.0967$ & $0.1798$ \\
$2461.767$ & $54949.5535$ & $1.7149$ & $9125.4479$ & $3.1523$ & $33.7601$ & $35.3355$ & $0.0987$ & $0.1807$ \\
$2462.793$ & $54945.9223$ & $1.6357$ & $9124.2665$ & $2.2858$ & $33.7449$ & $27.4409$ & $0.0952$ & $0.1812$ \\
$2463.802$ & $54948.8167$ & $1.8716$ & $9122.9704$ & $3.3019$ & $33.7807$ & $45.4873$ & $0.0962$ & $0.1752$ \\
$2464.744$ & $54946.3708$ & $1.6013$ & $9111.8036$ & $2.6251$ & $33.8372$ & $31.8717$ & $0.0916$ & $0.1789$ \\
$2467.665$ & $54939.5406$ & $2.5859$ & $9114.6408$ & $1.4407$ & $33.7794$ & $33.4307$ & $0.0974$ & $0.1811$ \\
$2471.654$ & $54941.8031$ & $1.9095$ & $9120.8601$ & $1.7179$ & $33.7775$ & $37.4413$ & $0.0975$ & $0.1815$ \\
$2473.559$ & $54941.1851$ & $1.5953$ & $9108.0420$ & $1.4481$ & $33.7786$ & $29.5206$ & $0.0994$ & $0.1817$ \\
$2473.609$ & $54944.5064$ & $1.5901$ & $9122.1321$ & $1.4729$ & $33.7577$ & $28.3784$ & $0.0995$ & $0.1818$ \\
$2474.613$ & $54944.8060$ & $2.3186$ & $9130.5444$ & $1.7984$ & $33.7461$ & $30.6980$ & $0.0982$ & $0.1825$ \\
$2475.583$ & $54940.0604$ & $1.8033$ & $9111.5859$ & $1.8403$ & $33.7598$ & $24.7552$ & $0.0992$ & $0.1776$ \\
$2476.602$ & $54938.6456$ & $2.2858$ & $9128.9792$ & $2.8436$ & $33.5903$ & $39.6079$ & $0.0982$ & $0.1829$ \\
$2477.579$ & $54935.1412$ & $3.3019$ & $9116.9772$ & $3.0458$ & $33.6390$ & $30.4956$ & $0.0962$ & $0.1800$ \\
$2477.694$ & $54933.4389$ & $2.6251$ & $9124.6195$ & $1.8033$ & $33.7387$ & $30.3916$ & $0.0986$ & $0.1819$ \\
$2478.739$ & $54939.0504$ & $3.1523$ & $9112.7792$ & $1.9095$ & $33.7201$ & $40.3934$ & $0.0980$ & $0.1817$ \\
$2479.654$ & $54941.7172$ & $1.6767$ & $9124.1326$ & $2.5859$ & $33.7251$ & $37.7955$ & $0.1023$ & $0.1830$ \\

\hline
\hline
\end{tabular}
\end{table*}

\end{document}